\documentclass[12pt]{article}  
 
\usepackage{epsfig} 
\usepackage{amsmath,amssymb,amsfonts} 
\usepackage{float,a4wide} 
\newcommand{\be}{\begin{equation}}  
\newcommand{\en}{\end{equation}} 
\newcommand{\bea}{\begin{eqnarray}} 
\newcommand{\ena}{\end{eqnarray}} 
 
\newcommand{\hbo}{\hbox to 1 true cm {\hfill } }  
\newcommand{\tr}{\hbox{tr}} 
\newcommand{\Tr}{\hbox{Tr}} 
  
\newcommand{\re}[1]{~(\ref{#1})}  
\newcommand{\fss}[1]{#1\!\!\!/} 
\newcommand{\Ds}{D\!\!\!\!/} 
\newcommand{\I}{\text{i}}  
\newcommand{\E}{\text{e}}  
\newcommand{\eff}{\text{eff}}  
\newcommand{\ferm}{\text{ferm}}  
\def\dslash{\partial\kern-.6em\slash} 
\def\kslash{k\kern-.5em\slash} 
\def\pslash{p\kern-.4em\slash} 
\def\Dslash{D\kern-.6em\slash} 
\def\Vslash{V\kern-.7em\slash} 
\def\vslash{v\kern-.5em\slash} 
\def\rslash{r\kern-.5em\slash} 
\def\qslash{q\kern-.5em\slash} 

\begin{document}  
\vglue 1truecm 
   
\vbox{ 
\hfill UNITU-THEP-18/02\\ 
$\text{}$\hfill  CERN-TH/2002-110  } 
\vbox{
\hfill 24 May 2002
} 
   
\vfil 
\centerline{\large\bf Fermion-induced quantum action of vortex systems }  
   
\bigskip 
\centerline{ Kurt Langfeld$^a$, Laurent Moyaerts$^{a}$,  
Holger Gies$^{b}$ } 
\vspace{1 true cm}  
\centerline{ $^a$ Institut f\"ur Theoretische Physik, Universit\"at  
   T\"ubingen } 
\centerline{D--72076 T\"ubingen, Germany} 
   
\bigskip 
\centerline{ $^b$ CERN, Theory Division } 
\centerline{CH-1211 Geneva 23,  Switzerland} 
 
\vfil 
\begin{abstract} 
\noindent
  The quantum action generated by fermions which are minimally coupled
  to abelian vortex background fields is studied in $D=2+1$ and
  $D=3+1$ Euclidean dimensions. We present a detailed analysis of
  single- and binary-vortex configurations using the recently
  developed method of worldline numerics. The dependence of the
  fermion-induced quantum action on the fermion mass and the magnetic
  fluxes carried by the vortices is studied, and the binary-vortex
  interaction is computed. Additionally, we discuss the chiral condensate
  generated from a dilute gas of vortices in the intermediate fermion
  mass range for the case $D=3+1$. As a byproduct, our findings
  provide insight into the validity limits of the derivative expansion,
  which is the standard analytical approach to inhomogeneous
  backgrounds. 
\end{abstract} 
 
\vfil 
\hrule width 5truecm 
\vskip .2truecm 
\begin{quote}  

PACS: 12.20.-m, 11.15.Ha

keywords: worldline, fermion-induced effective action, vortex,
Monte-Carlo simulation.
 
\end{quote} 
\eject 

\section{Introduction \hfill }  
 
The dynamics of surfaces plays an important role in many fields of
physics ranging from solid state physics and chiral quantum field
theories~\cite{Kaplan:1992bt,Kikukawa:2000dk,Chen:1999ne} to string
theory~\cite{Rubakov:2001kp}. The so-called vortices,
$d-2$-dimensional surfaces of the $d$-dimensional space, are of
particular interest in view of their phenomenological importance: in
solid state physics, freely moving vortices give rise to a
non-vanishing conductance of high $T_c$-superconductors, thereby
limiting their technical applicability~\cite{ptoday}. In the context
of Yang-Mills theory, it has recently been observed in lattice gauge
simulations that, in the continuum limit~\cite{Langfeld:1998jx}, the
Yang-Mills vacuum is populated by center vortices the core of which
can be detected in a certain gauge by a projection
technique~\cite{DelDebbio:1997xm, DelDebbio:1998uu}. A random gas of
these vortices can grasp the essence of quark confinement at zero
temperature, and the deconfinement phase transition at finite
temperatures can be understood as a vortex de-percolation
transition~\cite{Langfeld:1999cz, Engelhardt:2000fd}. Although
rigorous estimates signal that the core size of the full
(unprojected) field configurations is spread over the whole
spacetime~\cite{Kovacs:2002db}, an effective vortex model supplemented
by a particular core size describes certain low-energy observables 
remarkably well~\cite{Engelhardt:2000wr,Engelhardt:2000wc}.
 
\vskip 0.0cm  
In many applications, the surfaces appear as classical background fields  
(solitons) of an underlying field theory. The surface properties as well  
as the surfaces' interactions are determined by thermal and quantum  
fluctuations, respectively~\cite{Graham:2002fi,Graham:2001dy}.  
In the context of the layered superconductors, the spectrum of the  
quasi-particles encodes phenomenologically relevant information of the  
vortex background~\cite{gyg91,scho95}. In the context of Yang-Mills  
theory, a deep knowledge of the interplay of quarks with vortex-like  
solitons would help to describe hadron properties in the vortex  
picture.  
 
\vskip 0.0cm The determination of the vortex free energy generically
requires the calculation of functional determinants where the vortex
profile enters as a background field. Whereas the eigenvalue problem
of the single-vortex case with a $\theta$-function profile can be
solved analytically \cite{Gornicki:kq}, a numerical treatment seems
inevitable for realistic profiles and multi-vortex configurations. One
possible technique aims at the numerical integration of a Klein-Gordon
or Dirac type of equation associated with the differential operator
under consideration. For example, in
~\cite{Graham:2002fi,Graham:2001dy}, this approach has been further
developed and the numerical burden has been reduced to a quantum
mechanical computation of bound-state energies and scattering phase
shifts; in particular, the usually delicate issue of renormalization
can advantageously be inherited from standard perturbation theory in
this approach. The disadvantage of this approach is that numerical
complications increase with the degree of complexity of the background
field; up to now, only highly symmetric backgrounds have been treated
(see \cite{Pasipoularides:2000gg} and \cite{Bordag:1998tg} for the
single-vortex case). In this work, we put forward a recently proposed
numerical technique~\cite{Gies:2001zp,Gies:2001tj} which is based on
the ``string-inspired'' worldline method~\cite{Bern:1990cu}. In this
formalism, the effective action (functional determinant) is
represented in terms of first-quantized particle path-integrals, which
have turned out to be highly convenient for analytical computations
involving constant background fields (see~\cite{Schubert:2001he} for a
review). In our numerical realization of this formalism, Monte-Carlo
techniques can be exploited to estimate expectation values of
background-field dependent operators with respect to scalable
worldline ensembles, so-called {\em loop clouds}. This worldline
numerical scheme is formulated in coordinate space, which facilitates
a transparent renormalization, since the divergencies are associated
with local operators in coordinate space and the required counterterms
can easily be read off. The most important advantage of this formalism
is marked by the fact that the numerical algorithm can be realized
without any reference to the specific choice of the background field.
In particular, a high degree of symmetry of the background is not
required at all; on the other hand, if a symmetry is present, it can,
of course, be exploited for reducing numerical efforts.  Most
recently, this technique was adapted for worldline loops on a cubic
lattice, most convenient in the case of lattice gauge
simulations~\cite{Schmidt:2002yd}. Therein, it was pointed out that a
random walk easily generates loop clouds with the appropriate measure
required for the worldline approach.
 
In this paper, we will explore the free energy of vortex-type 
solitons, being supported by a $U(1)$ gauge field which is minimally
coupled to a fermion.  We will discuss the case of $2+1$ and $3+1$
spacetime dimensions and restrict ourselves to vortex
configurations which are static with respect to one or two directions
of spacetime, respectively.  The free energy will be studied for the
single and the binary-vortex configuration. In the latter case, we will
obtain the fermion-induced vortex-vortex interaction.
 
\vskip 0.0cm The paper is organized as follows: in the next section,
the worldline approach to the fermion determinant is introduced and
the renormalization procedure is discussed in detail for the case
$D=3+1$. The vortex configurations under investigation are introduced
in section \ref{sec:inter}. In section \ref{sec:one}, our numerical
result for the single-vortex configuration is presented and compared
with the estimate obtained from existing analytical studies. The
binary-vortex interaction is presented in section \ref{sec:bin}.
Fermion condensation due to a dilute gas of vortices is addressed in
section \ref{sec:cond}. Conclusions are left to the final section.

\section{Fermion determinants in the worldline approach}  
 
\subsection{Setup}  
\label{sec:setup}
 
In this paper, we shall investigate the case of one flavor of a
4-component Dirac fermion in $D=2+1$ and $D=3+1$ spacetime dimensions.
The fermion field is minimally coupled to a $U(1)$ gauge field
$A_\mu(x)$, which is considered as a background field.  The
fermion-induced effective action $\Gamma_{\ferm}$, which is a
free-energy functional, determines the probabilistic weight $\exp \{-
\Gamma _{\ferm} \}$. It is given by
\begin{equation}  
\Gamma_{\ferm} \;  
= \; - \, \ln \det (m-\I \Ds ) \; , \hbo \Ds \, := \, \fss{\partial} 
\, + \, \I \, \fss{A}   , 
\label{eq:1.1}  
\end{equation} 
where $m$ is the fermion mass, $\fss{A} = A_\mu (x) \, \gamma _\mu $, 
and $\gamma _\mu $ are the anti-hermitean Euclidean $\gamma$-matrices, 
$\gamma _\mu ^\dagger = - \gamma _\mu $. As a consequence,  
$\Ds$ is hermitean.  
 
\vskip 0.0cm  
Since the Dirac operator exhibits no spectral asymmetry in the 
4-component formulation, the effective action is real, and 
we obtain in Schwinger propertime regularization  
\begin{equation}  
\Gamma _{\ferm} \; = \;  \frac{1}{2} \int_{1/\Lambda^2}^\infty  
\frac{dT}{T}\, \E^{-m^2  T} \,  
\Tr \exp \Bigl\{ -  \Ds{}^{\,2} \,T\Bigr\}  \; ,  
\label{eq:1.2}  
\end{equation}  
where the scale $\Lambda $ acts as a UV regulator.  The idea of the 
worldline approach consists of rewriting  Eq.~(\ref{eq:1.2}) in terms 
of 1-dimensional path integrals (for a comprehensive review, see 
\cite{Schubert:2001he}). 
%
The worldline representation of  Eq.~(\ref{eq:1.2}) is given by  
\begin{equation}  
\Gamma _{\ferm} \; = \; \frac{1}{2} \,  
\frac{1}{(4\pi)^{D/2}} \int d^Dx_0 
\int_{1/\Lambda^2}^\infty  \frac{dT}{T^{(D/2)+1}} \; \E^{-m^2 T} \;  
4 \, \biggl\langle W_{\text{spin}}[A] \, - \, 1 \biggr\rangle_x,   
\label{eq:1.3}  
\end{equation}  
where a gauge-field independent constant has been dropped, and we 
have defined the spinorial counter gauge factor of the Wilson 
loop including the Pauli term (spin-field coupling),
\begin{equation}  
W_{\text{spin}}[A] \; = \; \frac{1}{4} \,  
\exp \biggl\{ i \int _0^T A_\mu (x) \; \dot{x}_\mu \; d\tau   
\biggr\} \; \; \;  
\tr \, P_T \, \exp\left(\frac{1}{2} 
 \int_0^T d\tau\, \sigma_{\mu\nu} F^{\mu\nu}\right).  
\label{eq:1.4}  
\end{equation}  
Here $ F^{\mu\nu} (x(\tau ))$ is the field strength tensor, and 
$\sigma _{\mu \nu } := i\, [\gamma _\mu, \gamma _\nu ]/2$ are 
hermitean Dirac algebra elements. The average $\langle \ldots 
\rangle_x $ in (\ref{eq:1.3}) is performed over an ensemble of closed 
worldlines, or {\em loop clouds}. A single loop is characterized by 
its worldline $x_\mu (\tau )$, $\tau \in [0,T]$ in $D$ dimensions. 
$P_T$ denotes path ordering with respect to the propertime $T$. The 
loops are centered upon a common average position, 
$$  
x_0^\mu \; := \; (1/T) \int_0^T \, d\tau \; x_\mu(\tau) \; ,  
\hbox to 5cm {\hfil (``center of mass'') } .  
$$  
The loop ensemble is generated according to the Gaussian weight: 
\begin{equation}
\exp \biggl[ - \frac{1}{4} \int_0^T d\tau \; \Bigl(\dot{x}_1^2 + 
\ldots + \dot{x}_D^2 \Bigr) \biggr] \; .  
\label{eq:gau}
\end{equation}
In practice, one greatly reduces the numerical work by generating a  
loop gas of unit propertime $T=1$ only, and by producing loop ensembles  
for a given propertime $T \not= 1$ by rescaling. This numerical method is  
discussed in some detail in~\cite{Gies:2001zp,Gies:2001tj}. 
 
\vskip 0.0cm 
Finally, we point out that the symmetry of the background field 
can easily be exploited to reduce the numerical burden. In our case, 
the background field is static with respect to $D-d$ space-time
dimensions, i.e., the gauge field depends only 
on the coordinates $x_1, \ldots, x_d$, $d<D$. Since the 
weight of the loop clouds is Gaussian, 
the loop average in $D$ dimensions can be evaluated by using 
$d<D$ dimensional loop clouds only, i.e., 
\begin{equation}
\biggl\langle F[A] \biggr\rangle_x \; = \; 
\frac{1}{\cal N} \; \int {\cal D}x_{1 \ldots d} \; F[A] \; 
e ^ { - \frac{1}{4} \int_0^T d\tau \; \Bigl(\dot{x}_1^2 + 
\ldots + \dot{x}_d^2 \Bigr) } \; , 
\end{equation}
where 
$$ 
{\cal N} \; = \; \int {\cal D}x_{1 \ldots d} \; 
e ^ { - \frac{1}{4} \int_0^T d\tau \; \Bigl(\dot{x}_1^2 + 
\ldots + \dot{x}_d^2 \Bigr) } \; .
$$
In particular, it is sufficient for the examples discussed below to
generate a single (large) 2-dimensional loop cloud to address the
properties of static vortices in $D=2+1$ and $D=3+1$, respectively.

\subsection{Renormalization}  
\label{sec:ren}
 
In the case of $D=2+1$ dimensions, the propertime integration in  
Eq.~(\ref{eq:1.3}) is finite at the lower bound. This implies that  
one might safely remove the regulator, i.e., $\Lambda \rightarrow  
\infty $, in this case. We therefore confine ourselves in this subsection  
to the case $D=3+1$, where the fermion-induced effective action discussed
above is actually the bare action and given by 
\begin{equation}  
\Gamma ^{\text{B}}_{\ferm} \; = \;  \frac{1}{2} \,  
\frac{4}{(4\pi)^{2}} \int d^4x_0 
\int_{1/\Lambda^2}^\infty  \frac{dT}{T^3} \; \E^{-m^2 T} \;  
\biggl\langle W_{\text{spin}}[A] \, - \, 1 \biggr\rangle_x \; .  
\label{eq:1.5}  
\end{equation}  
For small values of the propertime $T$, the expectation value $\langle 
W_{\text{spin}}[A] \, - \, 1 \rangle_x$ can be calculated analytically 
for the case of a loop cloud with $x_0$ being the center of mass.  One 
finds  
\begin{equation}  
\biggl\langle W_{\text{spin}}[A] \biggr\rangle_x \; = \; 1 \; + \; 
  \frac{1}{6} \, F_{\mu \nu }(x_0) F_{\mu \nu } (x_0) \; T^2 \; + \; 
\ldots \; , 
\label{eq:1.6}  
\end{equation}  
where the ellipsis denotes higher-order invariants that can be formed 
out of the field strength tensor, its dual, and derivatives thereof. 
Rewriting (\ref{eq:1.5}) as  
\begin{eqnarray} \Gamma ^{\text{B}}_{\ferm} &=& \; 
\frac{1}{8\pi^{2}} \int d^4x_0 \int_{1/\Lambda^2}^\infty 
\frac{dT}{T^3} \; \E^{-m^2 T} \; \biggl\langle W_{\text{spin}}[A] \, - 
\, 1 \, - \, \frac{1}{6} F^2(x_0) T^2 \, \biggr\rangle_x 
\label{eq:1.7} \\ 
&&+ \frac{1}{6} \, \frac{1}{8\pi^{2}}  
\int_{1/\Lambda^2}^\infty  \frac{dT}{T} \; \E^{-m^2 T} \;  
\int d^4x_0 \; F^2(x_0) \; ,  
\label{eq:1.8}  
\end{eqnarray}  
we observe that the part (\ref{eq:1.7}) of $\Gamma _{\ferm}^{\text{B}}$  
is finite if the regulator is removed, $\Lambda \rightarrow \infty $.  
Using  
\begin{equation}  
\int_{1/\Lambda^2}^\infty  \frac{dT}{T} \; \E^{-m^2 T} \; = \;  
- \; \ln \left( \frac{m^2}{\Lambda ^2} \right) \; - \; \gamma_{\text{E}} \;  
\; + \; {\cal O} \left( \frac{m^2}{\Lambda ^2} \right) \; , 
\label{eq:1.10}  
\end{equation}  
($\gamma_{\text{E}}$ is the Euler constant), we neglect irrelevant terms which  
are suppressed by powers of $1/\Lambda ^2$ in part (\ref{eq:1.7})  
and (\ref{eq:1.8}) and obtain  
\begin{eqnarray}  
\Gamma ^{\text{B}}_{\ferm} &=& \;  
\frac{1}{8\pi^{2}} \int d^4x_0 
\int_{0}^\infty  \frac{dT}{T^3} \; \E^{-m^2 T} \;  
\biggl\langle W_{\text{spin}}[A] \, - \, 1 \, - \, \frac{1}{6} 
F^2(x_0) T^2 \, \biggr\rangle_x  
\label{eq:1.11} \\ 
&&- \frac{1}{48 \pi^{2}}  
\left[\ln \left( \frac{m^2}{\Lambda ^2} \right) \; + \; \gamma_{\text{E}} 
\right] \; \int d^4x_0 \; F^2(x_0) \; .  
\label{eq:1.12}  
\end{eqnarray}  
The renormalized effective action $\Gamma _{\eff} $ is obtained by adding  
the bare ``classical'' action 
\begin{equation}  
\Gamma _{\eff} \; = \;  \Gamma ^{\text{B}}_{\ferm} \; + \; 
\frac{1}{4g_{\text{B}}^2} \int d^4x_0 \; F^2(x_0) \; ,  
\label{eq:1.13}  
\end{equation}  
where $g_{\text{B}}^2$ is the bare coupling constant,  
and we enforce the renormalization condition  
\begin{equation}  
\frac{1}{g_{\text{B}}^2(\Lambda ) } \; + \; \frac{1}{12 \pi^{2}}  
\left[ \ln \left( \frac{\Lambda ^2}{ \mu ^2} \right) - 
  \gamma_{\text{E}} \right]  
\; = \; \frac{1}{g_{\text{R}}^2 (\mu ) } \; .  
\label{eq:1.14} 
\end{equation}  
Here $g_{\text{R}} (\mu )$ denotes the renormalized coupling  
at a given renormalization point $\mu $, and we rediscover the QED 
$\beta$ function $\beta(g_{\text{R}}^2)=\mu\partial_\mu 
g_{\text{R}}^2(\mu)=g_{\text{R}}^4/(6\pi^2)$. Inserting (\ref{eq:1.11})  
and (\ref{eq:1.14}) into (\ref{eq:1.13}), the renormalized effective  
action is given by  
\begin{eqnarray}  
\Gamma _{\eff} &=& \Gamma_{F^2}+\Gamma_{\ferm},\nonumber\\  
\Gamma_{\ferm}&=&\frac{1}{8\pi^{2}} \int d^4x_0 
\int_{0}^\infty  \frac{dT}{T^3} \; \E^{-m^2 T} \;  
\biggl\langle W_{\text{spin}}[A] \, - \, 1 \, - \, \frac{1}{6} 
F^2(x_0) T^2 \,  \biggr\rangle_x,  
\label{eq:1.15} \\ 
\Gamma_{F^2}&=& \frac{1}{4} \left[ \frac{1}{g_{\text{R}}^2(\mu )} \, - \,  
\frac{1}{12 \pi^{2}} \, \ln \left( \frac{m^2}{\mu ^2} \right)\right] \;  
\int d^4x_0 \; F^2(x_0) \; .  
\label{eq:1.16}  
\end{eqnarray}  
In these equations, $\Gamma_{F^2}$ denotes the renormalized Maxwell
action; here we can impose fermion-mass-shell renormalization by
choosing $\mu=m$, so that the $\log$ term in (\ref{eq:1.16}) drops
out. In addition to the (trivial) classical term (\ref{eq:1.16}),
worldline numerics provides us with an explicit answer for the
renormalized fermion-induced quantum contribution $\Gamma_{\ferm}=\int
d^Dx\,{\cal L}_{\ferm}$ given by (\ref{eq:1.15}).
 
The renormalization scheme employed here can easily be related to
standard renormalization prescriptions for Feynman amplitudes. The
latter are formulated, for instance, by considering the renormalized
photon propagator $D_{\mu \nu }(p) = P_{\mu \nu } D(p^2)$ in the
Landau gauge, where $P_{\mu \nu }$ is the transverse projector, and
specifying a number $R$,
\begin{equation}
R := \frac{\partial }{ \partial p^2 } D^{-1} (p^2) \; \vert _{p^2=0} \; , 
\label{eq:i1} 
\end{equation}
which determines the wave function renormalization constant in a
particular renormalization scheme. Since the effective
action $\Gamma_{\eff}$ given above represents nothing but the
generating functional for one-particle irreducible Green's functions,
it is related to the renormalized photon propagator by
\begin{equation}
\Bigl( D^{-1} \Bigr) _{\mu \nu } (x,y) \; = \; 
\frac{ \delta ^2 \Gamma_{\eff} [A] }{ \delta A_{\mu }
(x) \; 
\delta A_{\nu } (y) } \; . 
\label{eq:i4} 
\end{equation}
We observe that the part (\ref{eq:1.15}) does not contribute to the
right-hand side of Eq.\re{eq:i1} in the limit $p^2 \rightarrow 0$;
this leads us to the desired result  
\begin{equation} 
R\; = \; 
\frac{\partial }{ \partial p^2 } D^{-1} (p^2) \; \vert _{p^2=0} \; =
\;  \frac{1}{g_{\text{R}}^2(\mu )} \, - \,  
\frac{1}{12 \pi^{2}} \,\ln \left( \frac{m^2}{\mu ^2} \right) \;
, 
\label{eq:i6} 
\end{equation} 
which determines the value of the coupling at any renormalization
scale $\mu$ for a given value of $R$. 
 
\section{ Vortex interfaces }  
\label{sec:inter}
 
The core of a vortex corresponds to a $D-2$ dimensional surface  
of $D$-dimensional Euclidean spacetime. The vortex field which we will  
discuss below is represented by a $U(1)$ gauge potential  
$A_\mu (x)$. In an idealized case, the gauge potential is  
singular at the vortex surface and can locally be represented by a  
pure gauge. Considering a closed line ${\cal C}$, the vortex can  
be characterized by the holonomy  
\begin{equation}  
\exp \biggl\{ i \oint _{\cal C} A_\mu (x) \; dx_\mu \biggr\} \; = \;  
\exp \{ i \, \phi \, L \} \; ,  
\label{eq:1.17}  
\end{equation}  
where $\phi $ is the magnetic flux carried by the surface, and  
$L$ is the linking number of ${\cal C}$ with the $D-2$ dimensional  
vortex surface. In the case $\phi = \pi $, the gauge potential describes  
a so-called center vortex in analogy to SU(2) Yang-Mills theory.  
 
\vskip 0.0cm In many physical applications (for instance, in the
context of center vortices in Yang-Mills theories) one observes that
the vortex surface possesses a finite thickness $d$ with respect to
the directions perpendicular to the flux, implying that the gauge
potential singularity in the surface is smeared.  We point out that
the physics of the extended object can be quite different from the
physics of the idealized vortex. In the latter case, one removes the
$D-2$ dimensional vortex surface from the base manifold. Quantum
fields are acting in the reduced coordinate space and obey certain
constraints at the singular subspace.  For purposes of illustration 
consider the case $\phi = 2 \pi $. One expects that the free energy of
the vortex is degenerated with the vacuum because the quantum fields
of both cases are related by a gauge transformation
(see~\cite{Diakonov:1999gg} for an illustration at the 1-loop level).
By contrast, in the case of the smeared vortex surface, the quantum
fields experience the magnetic flux of $\phi = 2 \pi $. Therefore one
does not expect a degeneracy of the smeared vortex configuration with
the vacuum.
 
\vskip 0.0cm  
In this first investigation, we will concentrate on plane  
vortex surfaces. For definiteness, we consider the gauge potential of
a vortex with flux $\phi = \varphi \pi $ and core size $d$ of the form
\begin{equation}  
A_\mu (x) \; = \; \frac{ \varphi }{2} \; \frac{1}{d^2 + x_1^2 +x_2^2 }  
\, (x_2,-x_1,0,...0)^T_\mu \; , \hbo \mu = 1 \ldots D \; .   
\label{eq:1.18}  
\end{equation}  
Since the worldline approach is manifestly gauge invariant, any other
choice of a gauge-equivalent configuration would give the same
fermionic interface energy.  The corresponding field strength is given
by
\begin{equation}  
F_{12} (x) \; = \; \frac{ \varphi \, d^2 }{ [d^2 + x_1^2 +x_2 ^2 ]^2 }  
\; .  
\label{eq:1.19}  
\end{equation}  
 
\vskip 0.0cm In the case of $D=2+1$ dimensions, the fermion-induced
effective action is proportional to the extent $L_t$ of the space-time
in time direction, 
\begin{equation}  
\Gamma ^{(3)}_{\ferm} \; = \; E \, L_t \; = \; L_t \; 2\pi \, 
\int _0^\infty  d\rho \; \rho \; {\cal L}_\mathrm{ferm}^{(3)} (\rho ) \; ,  
\label{eq:1.20}  
\end{equation}  
where $\rho $ is the radial coordinate of the $xy$-plane.  
$E$ can be interpreted as the energy of the static vortex  
soliton.  ${\cal L}_\mathrm{ferm}^{(3)} (\rho )$ is a radial 
energy density, i.e., the amount of energy which is stored in a 
cylindric shell $[\rho, \, \rho + d\rho ]$. 
In the case of $D=3+1$ dimensions and at a given time slice, the  
vortex core is given by a straight line of length $L_z$. The effective  
action is proportional to the extent $L_t \, L_z$, 
\begin{equation}  
\Gamma ^{(4)}_{\ferm} \; = \; \chi \, L_t \, L_z \; = \;  
 L_t \, L_z \; 2 \pi \, \int _0^\infty d\rho \; 
\rho \; {\cal L}_\mathrm{ferm}^{(4)} (\rho ) \; ,  
\label{eq:1.21}  
\end{equation}  
where $\chi $ is the string tension of the vortex line in the  
3D hypercube.

\section{ Quantum energy of the single-vortex configuration } 
\label{sec:one}

\subsection{ The derivative expansion}  

The standard analytical approach to effective actions and quantum
energies for nonhomogeneous backgrounds is the derivative expansion.
Here, the desired answer is expanded in terms of a small parameter
constructed from derivatives of the background field. In the present
case, there are two options for the expansion parameter, which can
symbolically be written as $\partial^2/m^2 \ll 1$ or $\partial^2/B(x)
\ll 1$, i.e., the derivatives of the background field should be
smaller than the fermion mass or the local field strength. It is
remarkable that closed-form expressions at the next-to-leading order
(NLO) level in $D=2+1$ and $D=3+1$ have been found in recent years
which are applicable in both cases; only one of the conditions
mentioned above has to be satisfied. The leading-order derivative
expansion, which agrees with the Euler-Heisenberg effective Lagrangian
for constant background fields, is given by
\cite{Heisenberg:1935qt,Schwinger:1951}
\begin{equation}
{\cal L}_{\ferm}^{d0}(x)=C_{d0}  \int _0^\infty \frac{dT}{T^\nu}\,
e^{-m^2\,T}\, [B(x)T\,\coth{(B(x)T)}-1+\{\text{c.t.}\}]. 
\label{eq:1.22}
\end{equation}
The next-to-leading (NLO) correction can be written as
\cite{Cangemi:1994by,Gusynin:1999}\footnote{A representation of the
  integral in terms of the Hurwitz Zeta function as well as asymptotic
  expansions have been found; see, for example, \cite{Gusynin:1999}.}
\begin{equation}
{\cal L}_{\ferm}^{d1}(x)=C_{d1} \frac{(\partial_i B(x))^2}{B^{\alpha} (x) } 
\int _0^\infty \frac{d\omega}{\omega^\rho}\,
e^{-\frac{m^2}{B(x)}\omega}\, \frac{d^3}{d\omega^3}[\omega\coth{(\omega)}].
\label{eq:1.23}
\end{equation}
The sum of both Lagrangians provides us with the NLO derivative
expansion of the fermion-induced effective Lagrangian. The prefactors
$C_{d0,1}$ and exponents $\alpha$, $\nu$ and $\rho$ depend on the
number $D$ of spacetime dimensions and are summarized in
Table~\ref{tab:prefactors}. Note that the counter term
$\{\text{c.t.}\}=-\frac{1}{3}(B(x)T)^2$ must be added to the integrand
in Eq.~(\ref{eq:1.22}) for $D$=3+1, as discussed in
Sect.~\ref{sec:ren} (cf. Eq.~(\ref{eq:1.7})); for $D=2+1$, there is no
counterterm, $\{\text{c.t.}\}=0$.

\vskip 0.0cm The expression (\ref{eq:1.23}) is valid for any value of
the ratio $m^2/B$, i.e., {\em either} $m^2$ {\em or} $B$ has to be
large compared with the inverse length scale squared set by the
variation of the background field. Near the vortex core, this length
scale is naturally given by the core size $d$. For a flux
$\varphi={\cal O}(1)$ considered in the following, we find from
Eq.~(\ref{eq:1.19}) that $B(x)\, d^2={\cal O}(1)$ near the core.
Simultaneously assuming small masses, $m\, d\lesssim 1$, neither of
the possible derivative expansion parameters is small. The NLO
Lagrangian reaches its validity limit, and one cannot expect reliable
results from the derivative expansion.  Below, it will turn out that
the quality of the NLO approximation strongly depends on the number of
spacetime dimensions. By contrast, for large masses $m d\gg 1$ or for
large radial distances from the vortex core $\rho \gg d$ , we expect
the NLO Lagrangian to give a reasonable answer, since one of the
possible expansion parameters is small.  Below, we will use the regime
of large masses (or large $\rho $) to gain insight into the range of
parameters to which worldline numerics is reliably applicable
(see~\cite{Gies:2001zp} and~\cite{Schmidt:2002yd} for a first
comparison of worldline numerics to the constant-field case).

\begin{table}[t]
\begin{center}
\label{tab:prefactors} 
\vskip 0.5cm
\begin{tabular}{ccc}
\ \hspace{1cm} & \hspace{1cm} $D$=2+1 \hspace{1cm} & \hspace{1cm} 
$D$=3+1 \hspace{1cm} \\ 
& & \\
\hline $C_{d0}$  & $\frac{1}{4\pi^{3/2}}$ & $\frac{1}{8\pi^2}$ \\
 $C_{d1}$ & $\frac{1}{4(4\pi)^{3/2}}$ & $\frac{1}{(8\pi)^2}$\\
$\alpha$ & 1.5 & 1 \\
$\nu$ & 2.5 & 3 \\
$\rho$ & 0.5 & 1 \\
\hline 
\end{tabular}
\caption{Prefactors and exponents of the derivative expansion.}
\end{center}
\end{table}

\subsection{ Worldline numerics: $D=2+1$ }  
Closely following the procedure outlined in~\cite{Gies:2001zp}, we
generate a loop ensemble with $n_e$ loops, where each loop is
represented by a set of $N$ spacetime points with coordinates $x_1
\ldots x_N$. The loop ensembles are generated with appropriate measure
(see discussion in subsection \ref{sec:setup}). We have used 10.000
and 20.000 dummy sweeps and found that for a proper thermalization, 
10.000 dummy sweeps are sufficient for the accuracy achieved in the
results shown below.  Furthermore, we have used $N \in {100,150,200}$
points specifying each loop. Whereas $N =100$ seems sufficient for our
purposes in the (2+1)-dimensional case, we take $N=200$ to generate
high-precision data for the case $D=3+1$.  Generically, we average
over $n_e=1000$ loop ensembles in the (2+1)-dimensional case, and
exceptionally use $n_e = 16000$ for certain applications in $D=3+1$.
Any dimensionful quantity quoted in the following is calculated in the
simulation in units of vortex thickness $d$. 

\begin{figure}[t]
\centerline{  
\epsfxsize=8cm 
\epsffile{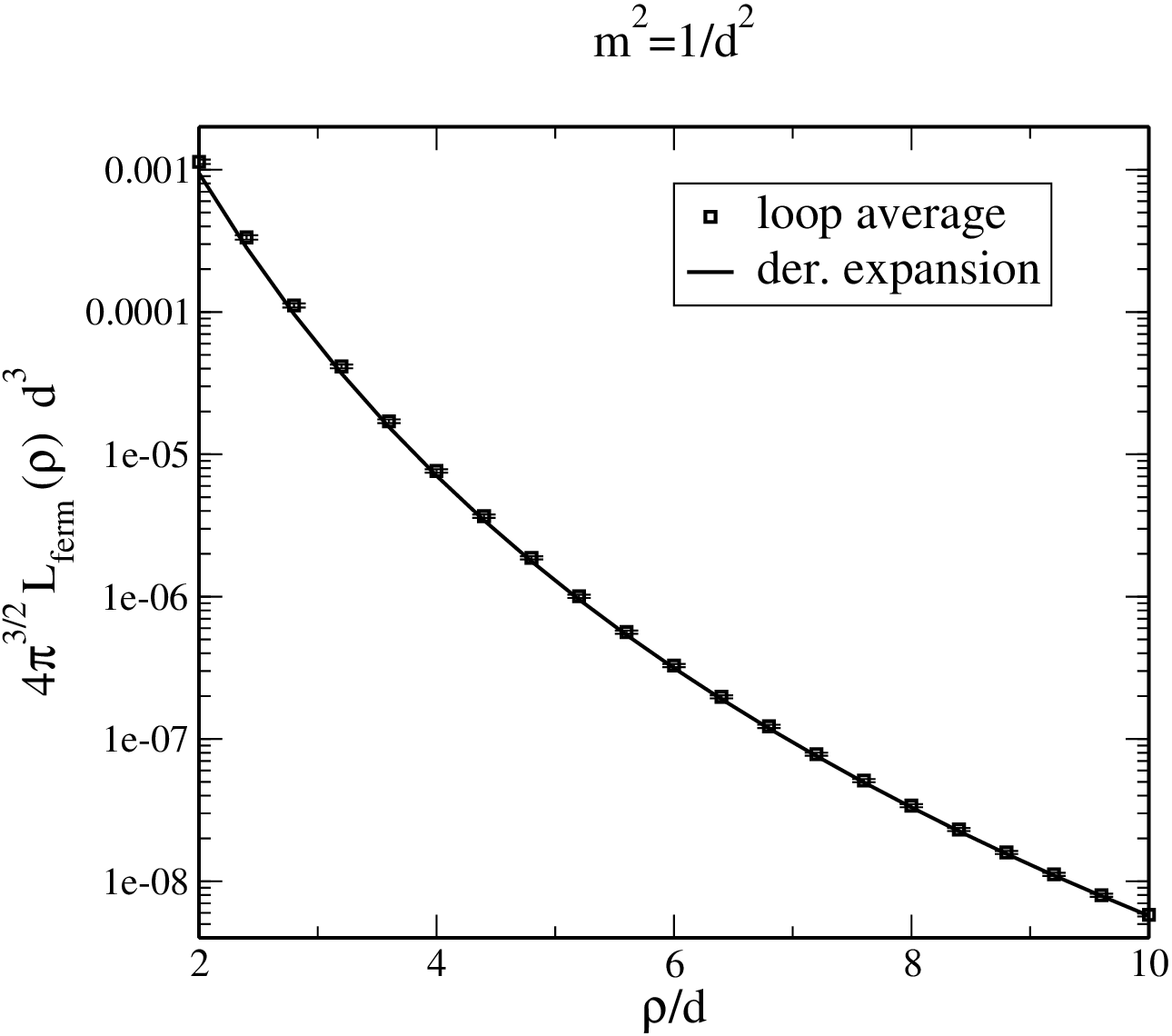} 
} 
\caption{Effective Lagrangian as a function of the radial distance
  $\rho$ to the vortex core for the case $m$=1,
  $\varphi$=1, $D$=2+1.  The NLO derivative expansion (solid line) is
  compared with the numerical computation (squares with error bars).}
\label{fig:L3log}
\end{figure}

\vskip 0.0cm 
In order to avoid a violation of gauge invariance
in the numerical computation, 
we have to deal with a subtlety concerning the
discretization of the integrals along the worldlines: a particular
loop in our ensemble is considered to be a polygon with
straight lines ${\cal C}_i$ connecting the points $x_i$ and $x_{i+1}$.  In
order to evaluate the holonomy in Eq.~(\ref{eq:1.4}), we consider the
``infinitesimal'' part
\begin{equation}  
\exp \biggl\{ i \int _{C_i} A_\mu (x) \; \dot{x}_\mu \; d\tau   
\biggr\} 
\label{eq:1.23a}  
\end{equation}
for each ${\cal C}_i$ separately and evaluate the integral
analytically using the vortex profile under
consideration. This procedure guarantees that our  
numerical result is invariant under gauge transformations of
the background gauge field $A_\mu (x)$ for any number $N$ of points 
defining the polygons. Moreover, the flux enclosed by the polygon is
exactly taken into account as desired (and not only within
discretization errors). 
Of course, this procedure can be generalized to arbitrary vortex
profiles for which the  ``infinitesimal'' integrations along the
${\cal C}_i$'s can be performed numerically; thereby, the properties
mentioned above can be preserved to any numerically desired
precision. However, we should stress that the use of a smooth gauge
(e.g., covariant gauges) for the background field is recommended in
this case; this facilitates a fast convergence of the numerical
integration. 

\begin{figure}[t]
\centerline{  
\epsfxsize=6cm 
\epsffile{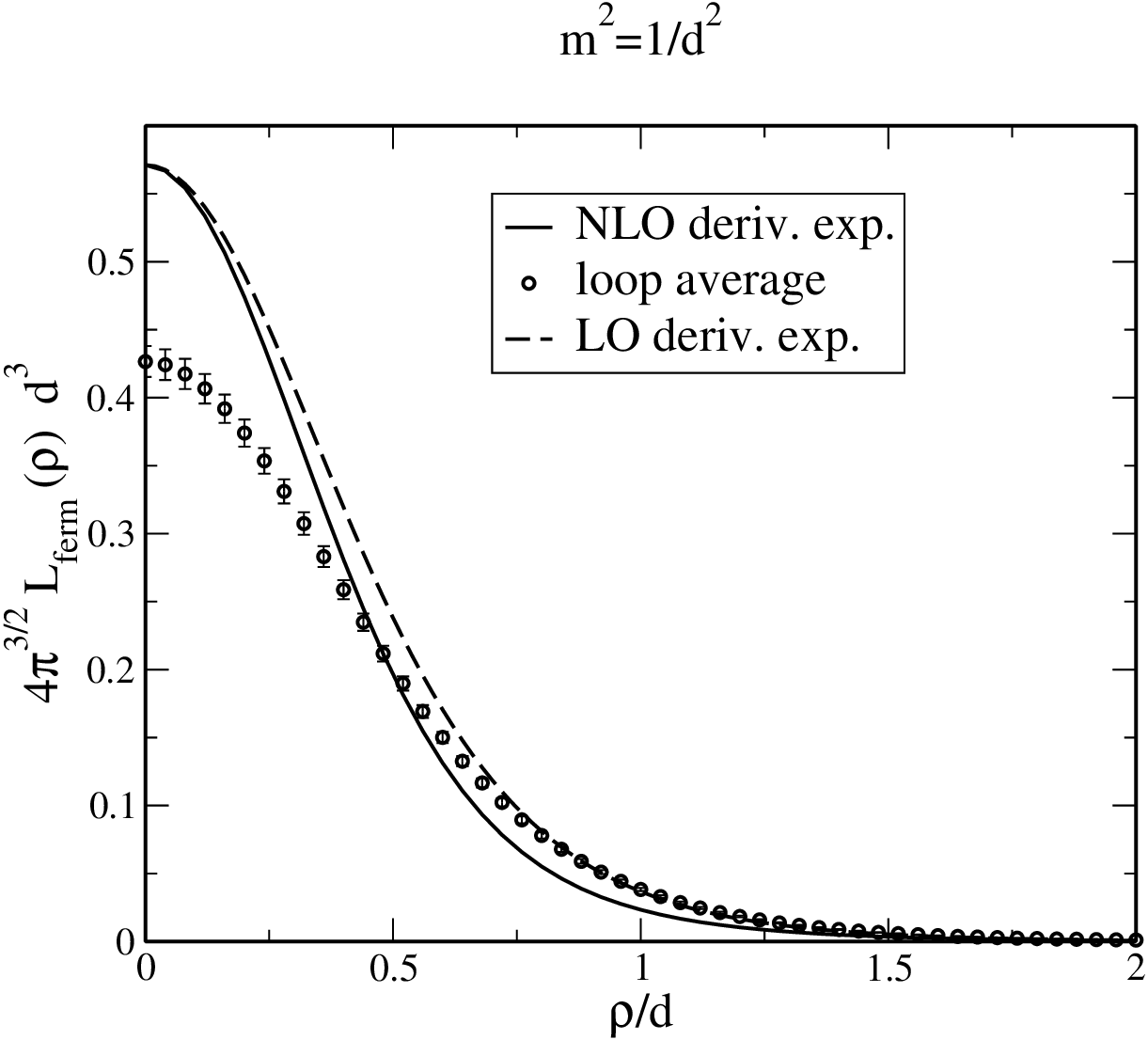} 
\epsfxsize=6cm 
\epsffile{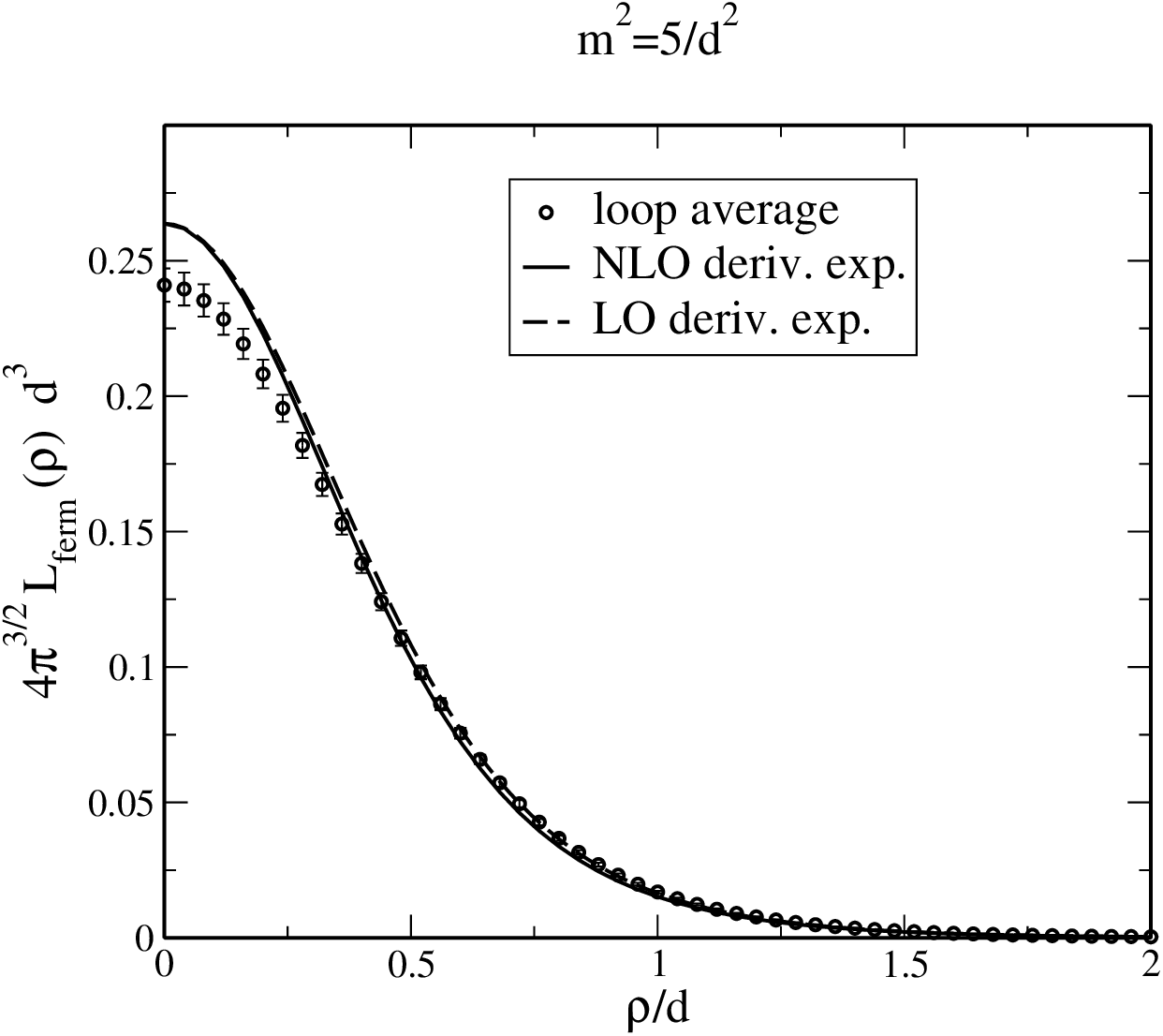} 
} 
\caption{Effective Lagrangian in the small $\rho$ region
  for the cases $m$=1 (left panel) and $m^2$=5 (right panel),
  $\varphi$=1, $D$=2+1.}
\label{fig:L3lin}
\end{figure}  
\vskip 0.0cm In order to test our numerical approach, we calculate the
effective Lagrangian ${\cal L}_\mathrm{\ferm}( \rho )$ for large
values of $\rho $, $\rho \gg d $, where the derivative expansion
is expected to give reliable results. The result of the numerical
worldline approach is compared with the derivative expansion in
Fig.~\ref{fig:L3log} for $D=2+1$. The agreement between the two curves
is satisfactory; in particular, the numerical approach is able to
compute ${\cal L}_{\ferm}$ over a range of many orders of magnitude.
In the region close to the core of the vortex, i.e., $\rho \approx d$,
the gradients of the background field are as large as the field
itself, so that the reliability of the derivative expansion now
  depends on the value of the mass. Our numerical findings for this
  regime are shown in Fig.~\ref{fig:L3lin} for $m^2=1$ (left panel)
  and $m^2=5$ (right panel). For larger masses, we observe a good
  qualitative and a reasonable quantitative agreement. But even for
  the small mass value $m^2=1$, there is at least qualitative
  agreement between the numerical result and the derivative expansion,
  indicating that the applicability of the derivative expansion can be
  pushed to its formal validity limit in $D=2+1$. This observation is
  also supported by the fact that the NLO term\re{eq:1.23} is only a
  small correction to the zeroth-order result\re{eq:1.22}. Moreover,
  we expect that the (up to now unknown) NNLO correction, which is
  sensitive to the curvature of the field strength, improves the
  result near the vortex core. In this sense, it is reassuring to
  observe that the numerical result agrees with the NLO derivative
  expansion precisely at the turning point of the curve at
  $\rho\simeq0.5$, because the NNLO correction must vanish here. We
  should finally stress that for even smaller masses $m<1$, the
  discrepancy between our numerical result and the derivative
  expansion increases, so that the derivative expansion should be
  abandoned here.

\begin{figure}[t] 
\centerline{  
\epsfxsize=7cm 
\epsffile{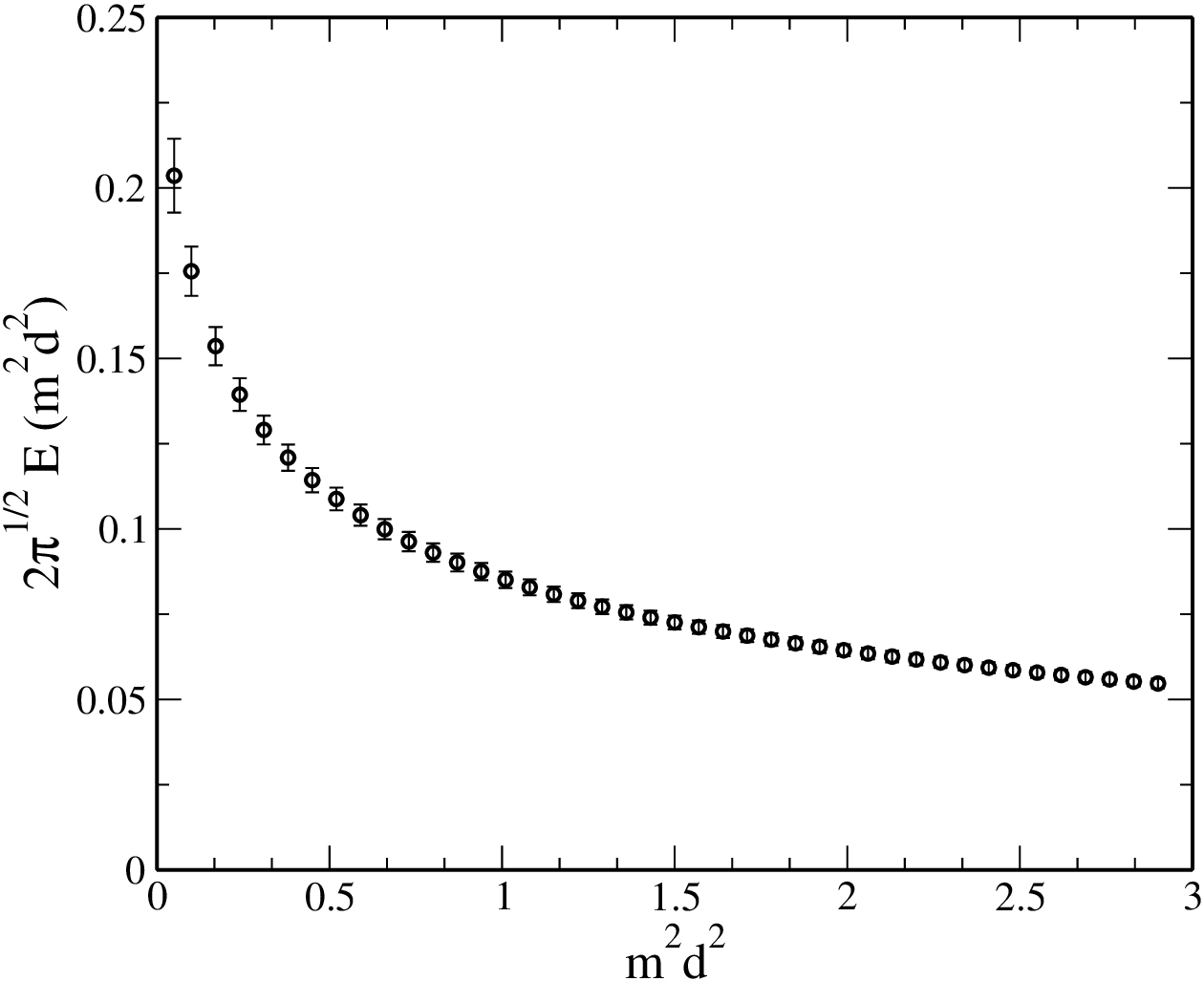} 
\hspace{0.2cm}
\epsfxsize=7cm 
\epsffile{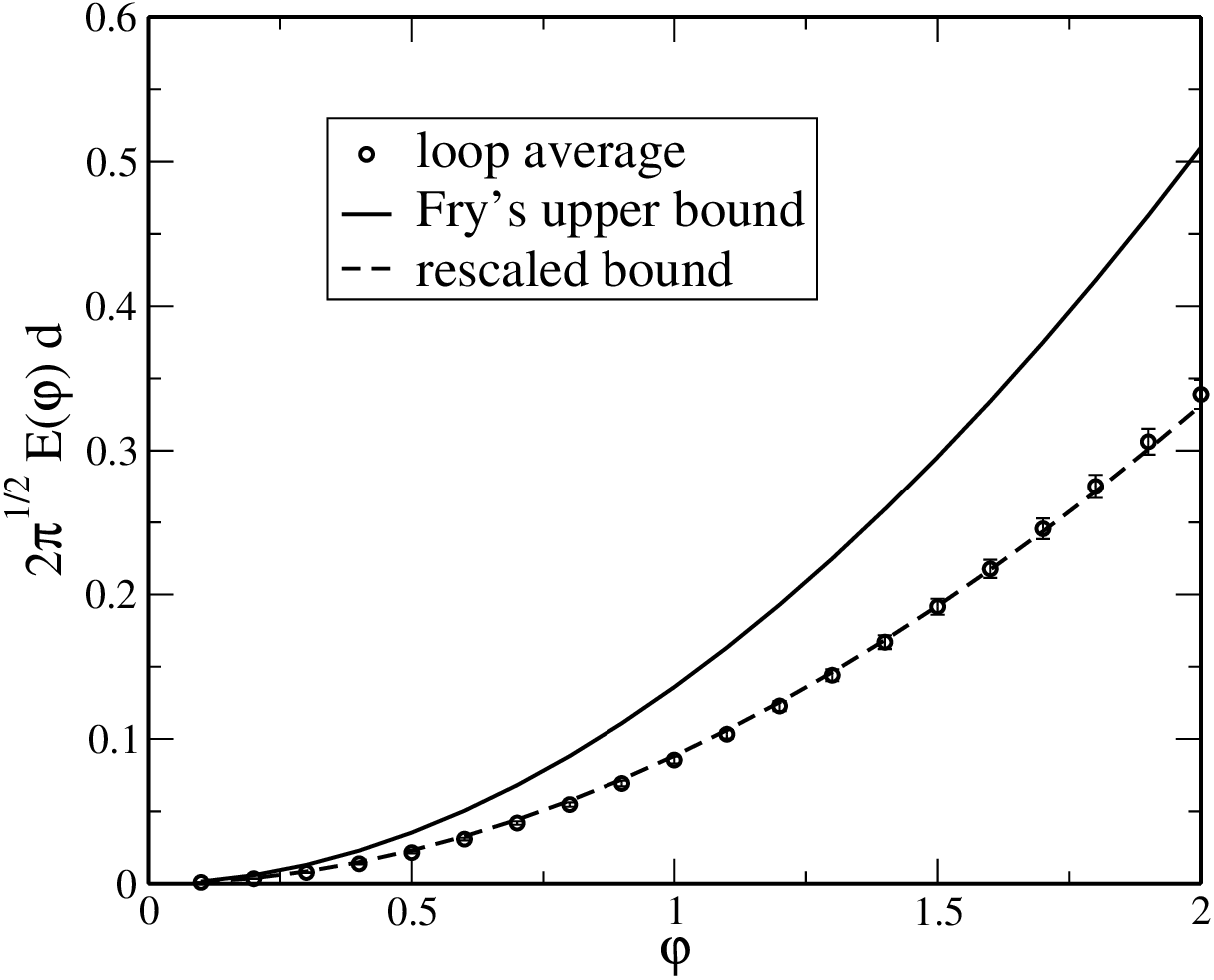} 
} 
\caption{Quantum energy as a function of the mass
  $m^2$ for $\varphi$=1 (left panel) and as a function of the flux
  $\varphi $ carried by the vortex for $m=1$ (right panel) in
  comparison with Fry's upper bound given in Eq.\re{eq:fry}.}
\label{fig:S3m} 
\end{figure} 
\vskip 0.0cm 
Let us now examine the fermion-induced quantum energy $E$
of the vortex soliton as defined in Eq.~(\ref{eq:1.20}),
\begin{equation}
E=\Gamma^{(3)}_{\ferm}/L_t=2\pi \int _0^\infty \! d\rho\,
\rho \,{\cal L}_{\ferm}(\rho),
\label{eq:1.24}
\end{equation}
which we obtain by numerically integrating the effective Lagrangian.
Our result for this energy is shown in Fig.~\ref{fig:S3m} as function
of the fermion mass $m$ in units of the vortex thickness $d$. Since
fermion fluctuations are suppressed with increasing mass, the quantum
energy decreases with increasing $m$, and vanishes in the large mass
limit. For phenomenological purposes, it is important to notice that
the quantum energy is positive. This implies that potential effective
models for vortex dynamics have to account for the fact that 
vortex nucleation is suppressed by the fermion-induced effective
action in $D=2+1$ dimensions. 

\vskip0.0cm Furthermore, let us consider the variation of the quantum
energy with respect to the flux $\phi$, which is carried by the
vortex.  Our numerical result is shown in Fig.~\ref{fig:S3m} (right
panel).  We find that the energy is monotonically increasing with the
flux $\phi $. For $\phi\equiv\varphi \pi$=2$\pi$, the vortex
configuration approaches asymptotically $(\rho \rightarrow \infty)$ a
pure gauge.  As in the instanton case, the energy is
nonvanishing due to the finite extension of the vortex core.  It is
interesting to compare our result for $E(\varphi)$ with a general
result for fermionic determinants described by M.~Fry in  
\cite{Fry:1996iq}; therein a lower bound has been derived for
unidirectional magnetic fields in $D=2+1$, which translates into an
upper bound $E_{\text{b}}$ for the quantum energy of our vortex
configuration given by
\bea 
E(\varphi) &\leq & E_{\text{b}}(\varphi),
\label{eq:fry} \\ 
E_{\text{b}}(\varphi) &=&
\left( \frac{1}{d}\right) \frac{1}{6} 
\Bigl[2-3 \varphi-2\sqrt{1+\varphi}+\varphi\sqrt{1+\varphi} 
      +3\sqrt{\varphi} \,\text{Arsinh} (\sqrt{\varphi})\Bigr] 
\nonumber 
\ena 
for the case $md=1$ and Dirac 4-component spinors. For other 
values of $md$, this formula receives a total factor
of $(md)^3$ and the flux has to be replaced by
$\varphi\to\varphi/(md)^2$. As shown in Fig.~\ref{fig:S3m}, our
numerical result lies well within this bound. More remarkable is the
fact that the functional dependence of our result agrees with the
bound within the error bars, if the bound is scaled by a factor of
roughly $0.65$. 


\vskip 0.0cm As a further check, we have compared all our
above-mentioned 
results with those of \cite{Pasipoularides:2000gg}, where the
single-vortex case with profile functions different from ours was
considered within the phase-shift approach. We find good agreement
within the error bars except for a global factor of 2 by which the
result of \cite{Pasipoularides:2000gg} for $E(\varphi)$ is larger.

\vskip 0.0cm Let us finally point to a problematic feature of our
present approach. Note that our statistical error bars increase in the
regime of small fermion masses in Fig.~\ref{fig:S3m} (left panel),
which has the following origin: the Pauli term in Eq.~(\ref{eq:1.4}),
$$
\tr \, P_T \, \exp\left(\frac{1}{2} 
 \int_0^T d\tau\, \sigma_{\mu\nu} F^{\mu\nu}\right) \; , 
$$ 
can favor large loops $(T\gg d^2)$ for the present unidirectional
magnetic field. On the other hand, the holonomy 
factor, 
$$
\exp \biggl\{ i \int _0^T A_\mu (x) \; \dot{x}_\mu \; d\tau   
\biggr\} \; , 
$$ 
changes its sign rapidly in this case when the loop under consideration
is slightly deformed. This implies that a finite
result in the small-mass regime arises after subtle cancellations, and 
is therefore hardly accessible to the present numerical
formulation. For finite values of the mass, potentially large Pauli
term contributions are suppressed by the factor $\exp(-m^2 T)$, which
solves the numerical problem for large loops (large $T$).\footnote{In
  the constant-field case, this cancellation problem already occurs 
  for $m^2\lesssim B$ as observed in \cite{Gies:2001tj}. In the
  present vortex case, we can afford much smaller masses. In general,
  the smaller the coherence length of the field, the less serious 
  the cancellation problem. The usually considered homogeneous and
  unidirectional field configurations are therefore rather
  pathological in this respect.}
However, this cancellation problem can become serious for the approach
to the chiral limit; here, a solution to the problem has to be
implemented in the algorithm on the analytical level rather than by
brute-force numerical means.

\subsection{ Worldline numerics: $D=3+1$ }  

\begin{figure}[t] 
\centerline{  
\epsfxsize=7cm 
\epsffile{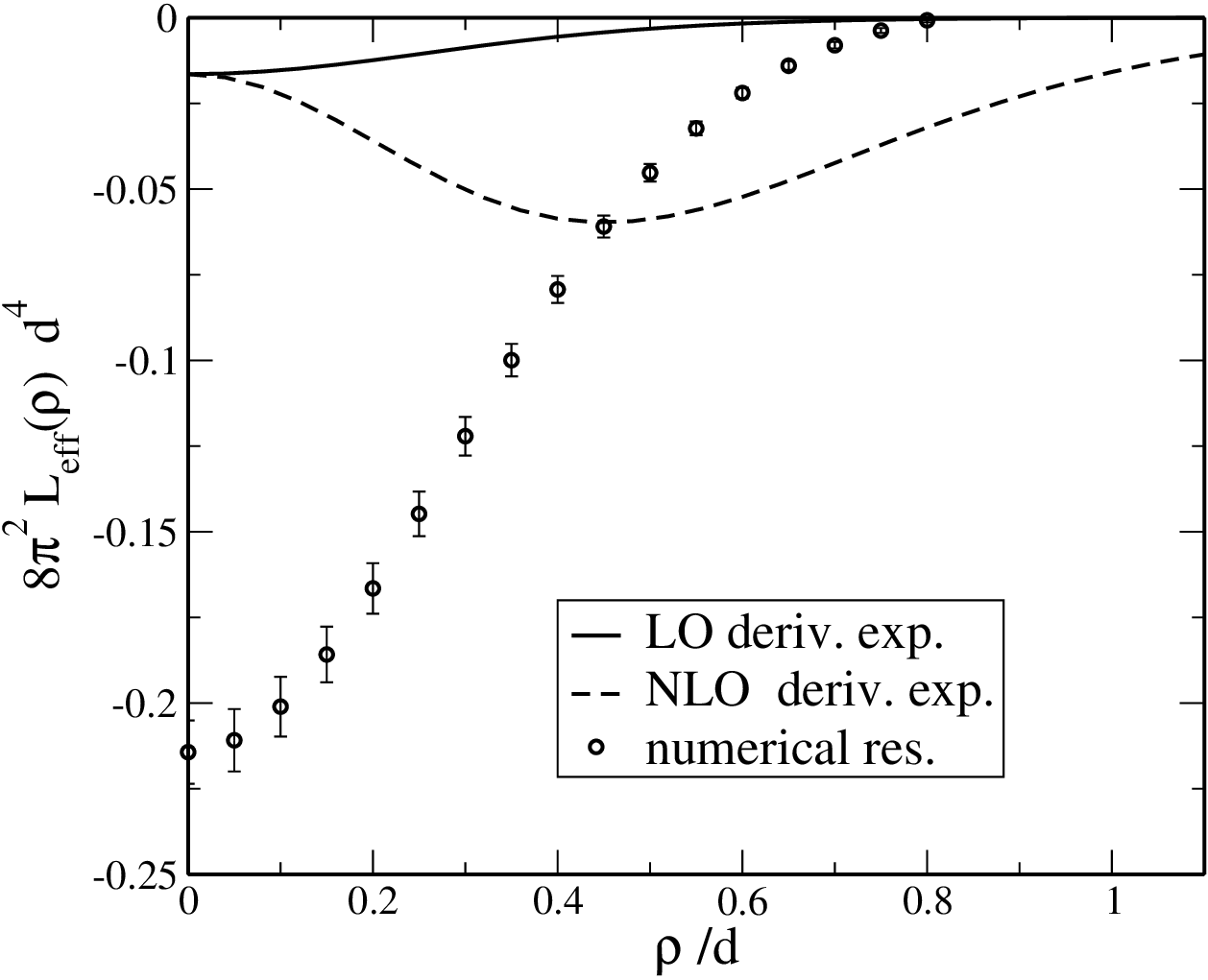} 
\epsfxsize=7cm 
\epsffile{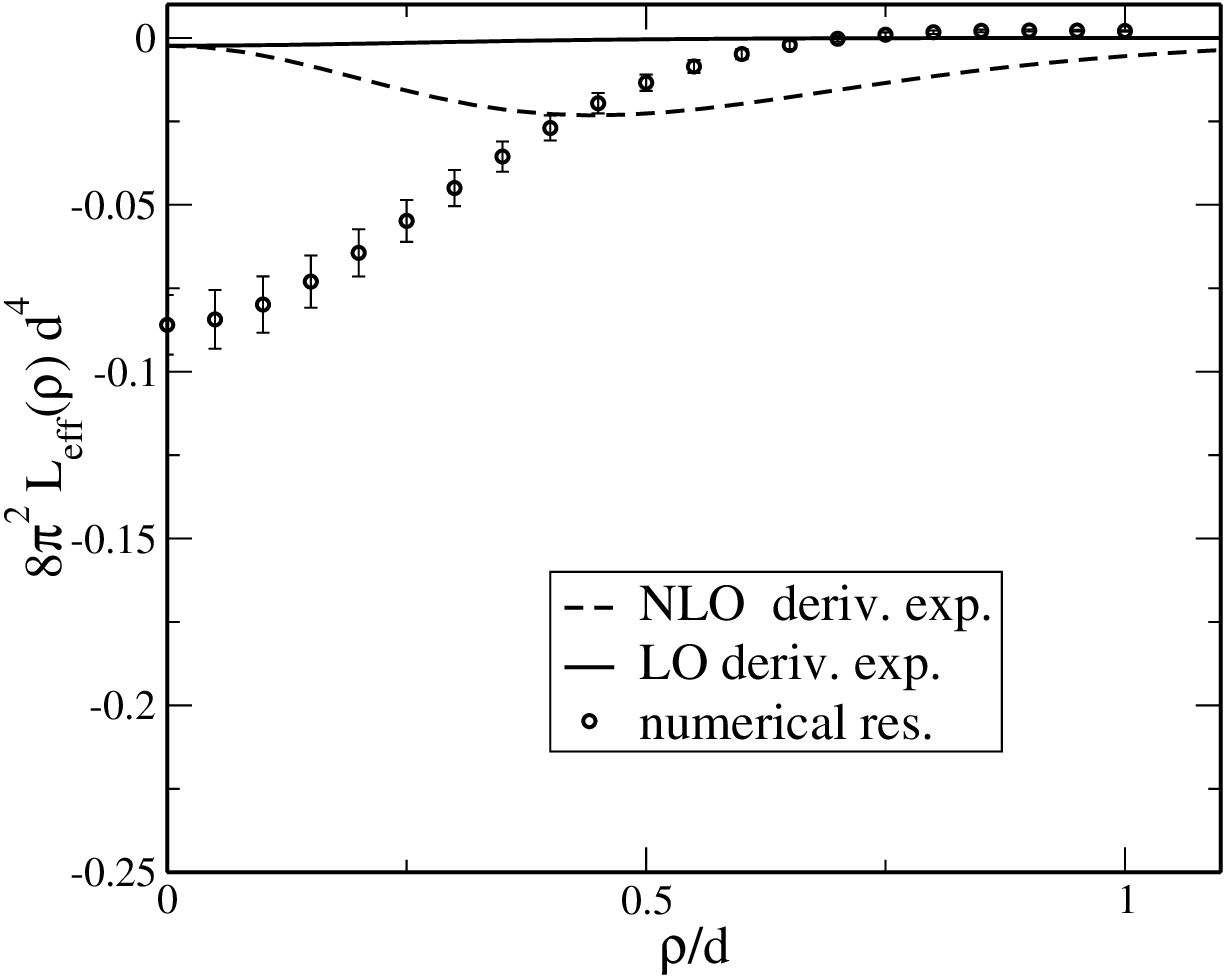} 
} 
\caption{Effective Lagrangian for $m^2=1$ (left panel) and $m^2=3$ 
   (right panel), $\varphi$=1, $D=3+1$. } 
\label{fig:L4lin}
\end{figure} 

Similarly to the previous studies of the $D=2+1$ case, we investigate
the effective Lagrangian ${\cal L}_\mathrm{ferm}$ as a function of $\rho
$ and compare the result with the one obtained from the derivative
expansion (\ref{eq:1.22},\ref{eq:1.23}) (see Fig.~\ref{fig:L4lin}).
Contrary to $D=2+1$, we observe that unless $\rho \gg d$, the leading
order (LO) of the derivative expansion falls far too short of
reproducing the 
numerical result for $m^2={\cal O}(1)$. For instance, for $m^2=1$, we
find an order-of-magnitude difference at $\rho=0$   
(Fig.~\ref{fig:L4lin} (left panel)).  Moreover, the NLO contribution
of the derivative expansion also exceeds the leading order result by
almost an order of magnitude for $\rho \approx d$. This signals the
break-down of the derivative expansion for the case of the field, its
gradient and the mass being all of the same order. Contrary to the
case of $D=2+1$, the applicability of the NLO derivative expansion
cannot be pushed to its formal validity limits $md \approx 1$.  Even
for larger values of the mass, $m^2=3$ (right panel), there is only a
minor improvement of the quality of the NLO derivative expansion. Of
course, the NNLO contribution could, in principle, improve the results
of the derivative expansion, but this would only emphasize the fact that
there is no clear hierarchy from term to term in the derivative
expansion.

\vspace{.0cm} We believe that the striking difference to the $D=2+1$
dimensional case is indeed remarkable and points to a deeper reason in
terms of a renormalization effect. To illustrate this, we note that
the quantity $\langle W_{\text{spin}}-1\rangle$ occurring in the
propertime integrand is positive for the vortex background, whereas
the counterterm $-\frac{1}{3} B^2(x) T^2$ is negative. Since the
effective Lagrangian is largely negative as seen in
Fig.~\ref{fig:L4lin}, it is mainly driven by the counterterm. Now the
leading-order derivative expansion obviously overestimates the value
of $\langle W_{\text{spin}}-1\rangle$ near the vortex core, since it
is a local expansion. The true value as seen in the numerical
computation is much smaller because it is a nonlocal average over the
extended {\em loop cloud} that also ``feels'' the much weaker field at
a radial distance from the core. The final value of the total
effective Lagrangian at a point $x$ therefore results from a
nontrivial interplay between nonlocal (and nonlinear) vacuum
polarization ($\sim\langle W_{\text{spin}}-1\rangle$) and a local
definition of the coupling giving rise to a local counterterm. In
regions where the background field varies rapidly, such as the near
vortex core in our case, this interplay can lead to an
order-of-magnitude enhancement of the effective Lagrangian as compared
with the constant-field approximation (leading-order derivative
expansion). In our opinion, this phenomenom clearly deserves further
investigation.
\begin{figure}[t]
\centerline{  
\epsfxsize=6cm 
\epsffile{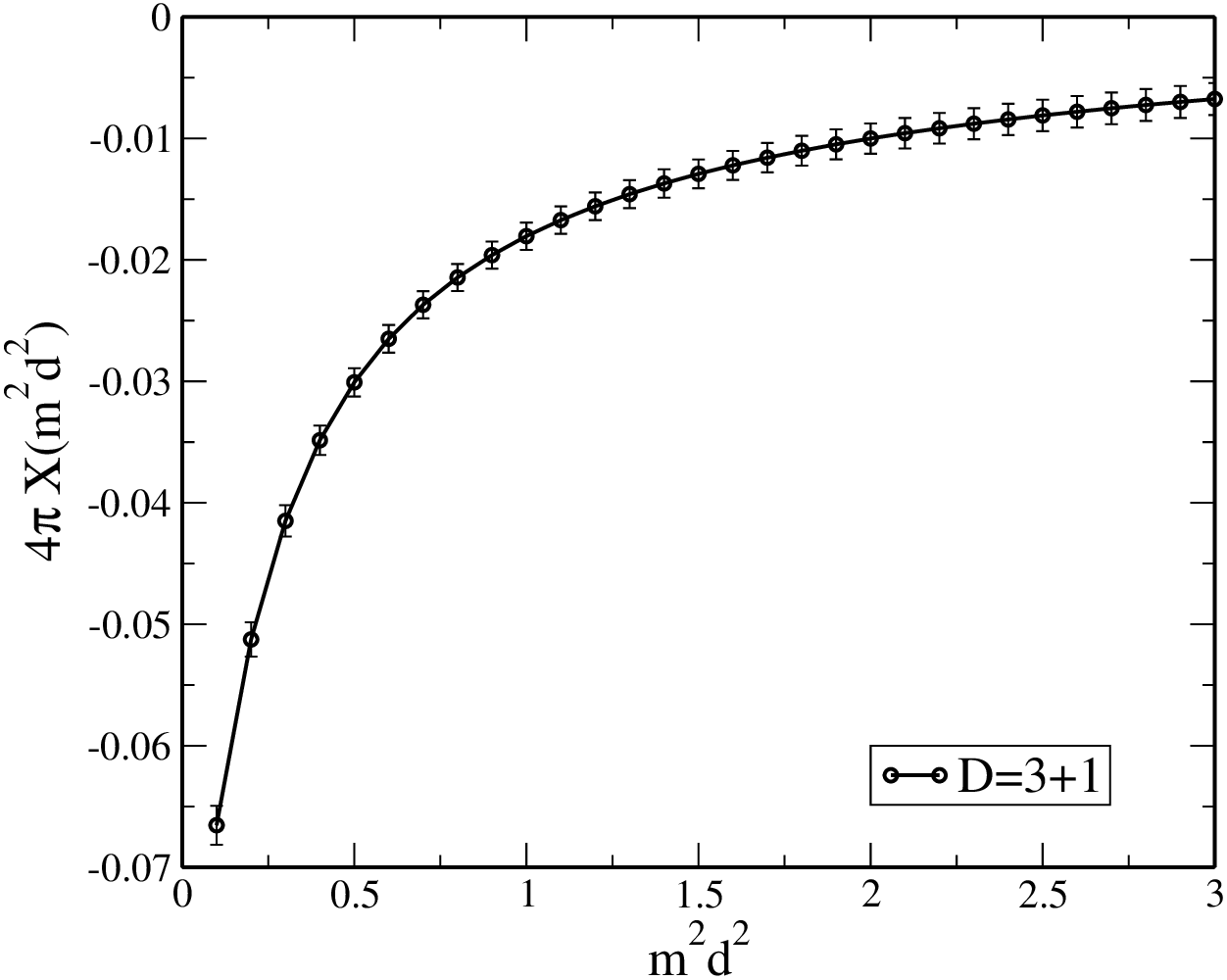} 
\epsfxsize=6cm 
\epsffile{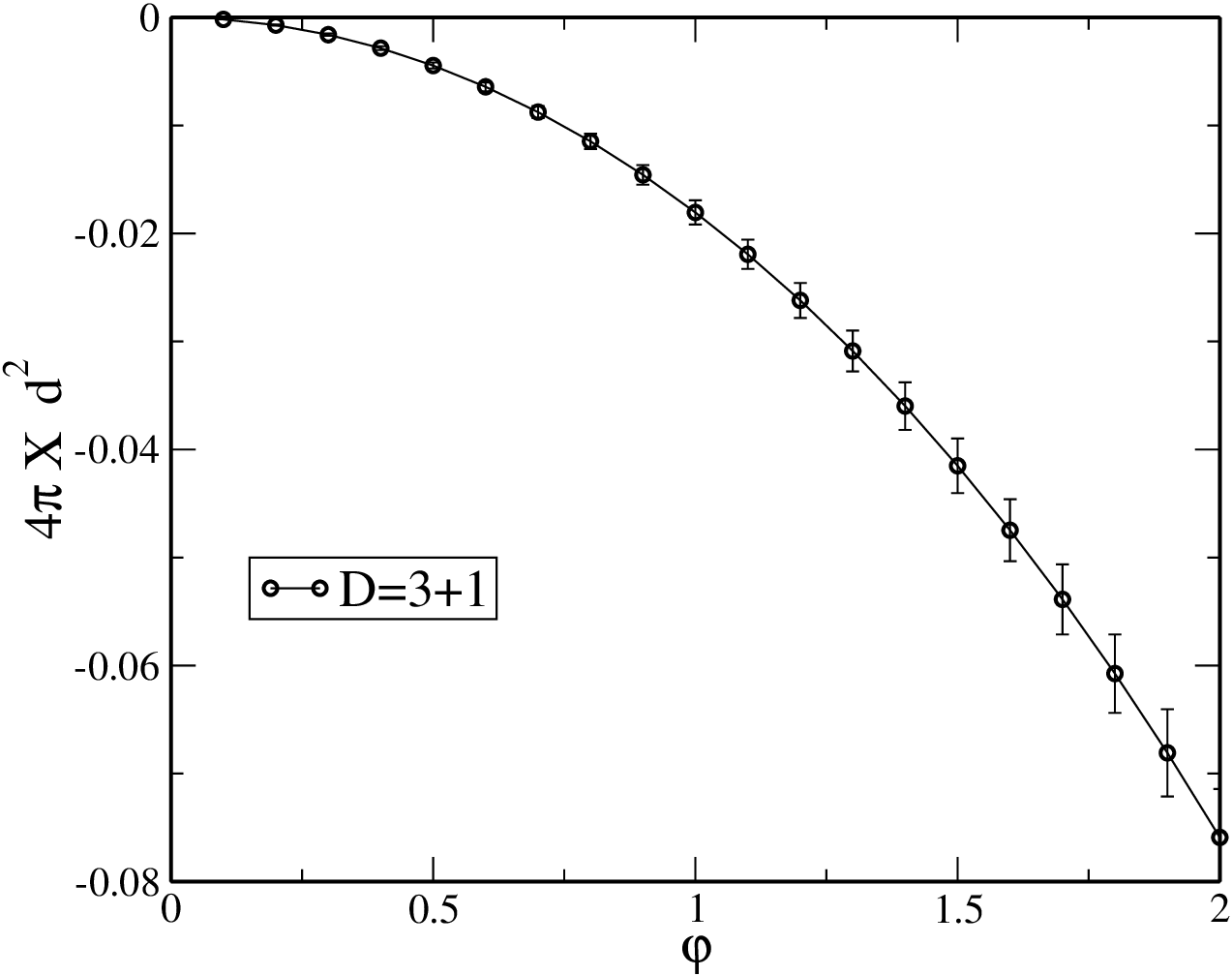} 
} 
\caption{Effective action as a function of the dimensionless quantity
$m^2d^2$, $\varphi$=1, $D$=3+1 (left panel), and as function of the 
flux, $m^2d^2=1$ (right panel).}
\label{fig:S4m}
\end{figure}
\vskip 0.0cm Returning to our numerical study of the one-vortex
background, we calculate the string tension $\chi$ as defined in
Eq.~(\ref{eq:1.21}) as a function of the fermion mass $m$ and plot it
in Fig.~\ref{fig:S4m} for $\varphi=1$. The negative values of $\chi $
show that the fermion-induced effective action $\Gamma_{\ferm}$ {\it
  favors } the nucleation of vortices. Since the modulus of this
effective action increases if the vortex thickness $d$ is decreased,
the fermionic part of the vortex action supports the existence of thin
vortices.  These results are in contrast to those of the case $D=2+1$, 
where the effective action turned out to be positive (see
Fig.~\ref{fig:S3m}). This sign difference is again related to
renormalization effects mediated by the counterterm.

\section{ Binary-vortex interactions } 

\label{sec:bin}

\begin{figure}[t]
\centerline{  
\epsfxsize=6.2cm 
\epsffile{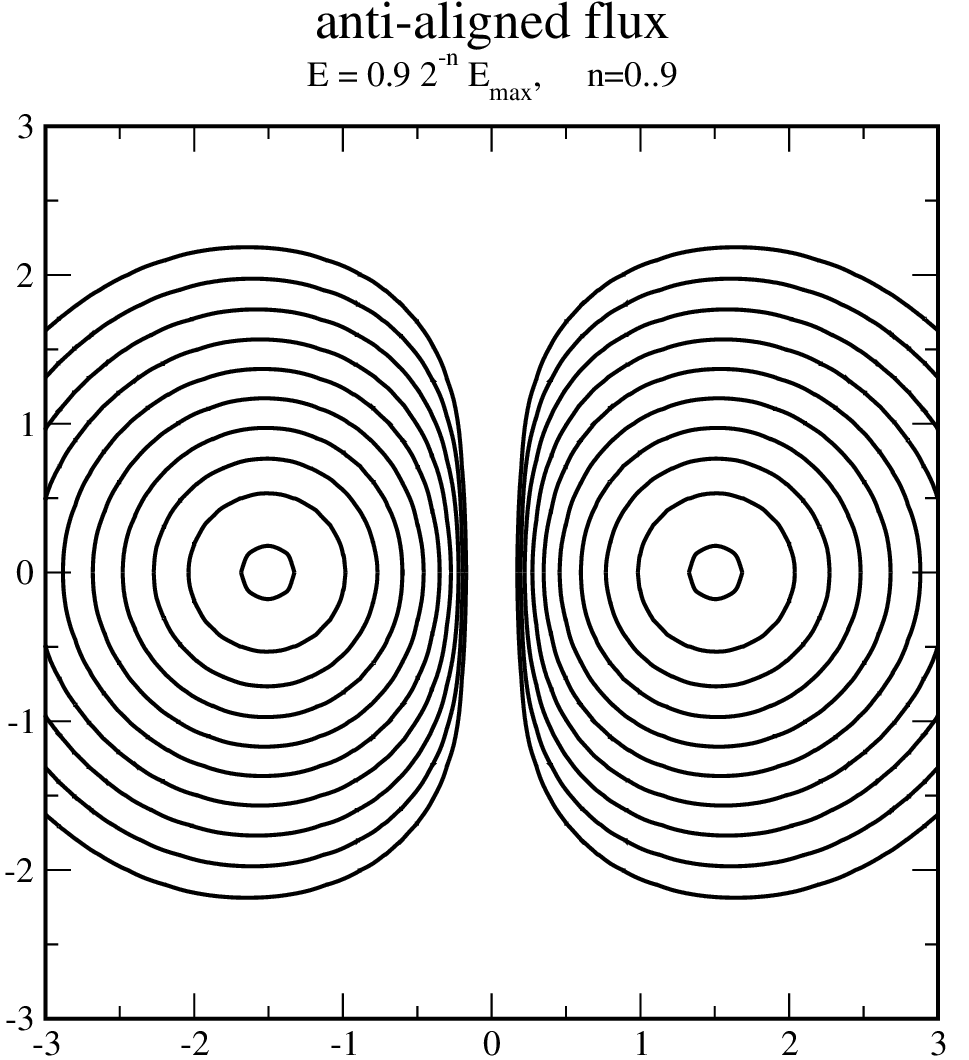} 
\hspace{0.5cm}
\epsfxsize=6cm 
\epsffile{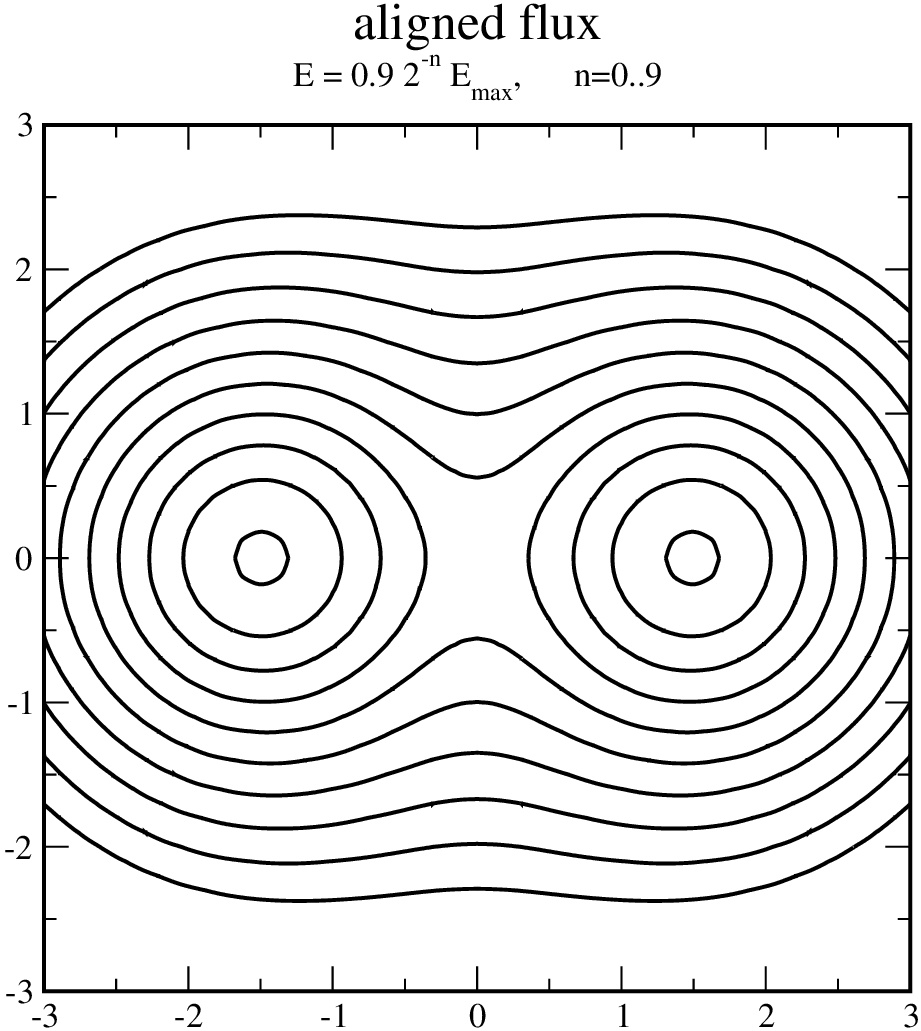} 
} 
\caption{ Effective Lagrangian $L_\mathrm{ferm}^{(3)}(x,y)$ as function
   of the $xy$-plane for the two vortex configuration: parallel 
   (right panel) and anti-parallel (left panel) orientation of the flux, 
   $m^2d^2 = 0.5$, $\varphi$=1, $D$=2+1.}
\label{fig:cont}
\end{figure}
Since we are investigating the case of Abelian gauge configurations,
the binary-vortex configuration is given by the superposition of 
two single vortex gauge fields $A_\mu (x)$ (\ref{eq:1.18}),
\begin{equation}
A_\mu ^{(2)}(x) \; = \; A_\mu \Bigl( x-\frac{l}{2} \Bigr) 
\; \pm \; A_\mu \Bigl( x+\frac{l}{2} \Bigr)  \; , 
\label{eq:5.1} 
\end{equation}
where $l$ denotes the vortex distance. Below, we will study the case 
of the so-called center vortices the flux of which is given by 
$\varphi = 1 $. The relative sign between the gauge fields on 
the right-hand side of Eq.~(\ref{eq:5.1}) corresponds to the relative 
orientation of the fluxes: the plus sign corresponds to an equal 
orientation of the flux in each vortex, while the minus sign 
signals an opposite orientation.

\begin{figure}[t]
\centerline{  
\epsfxsize=8cm 
\epsffile{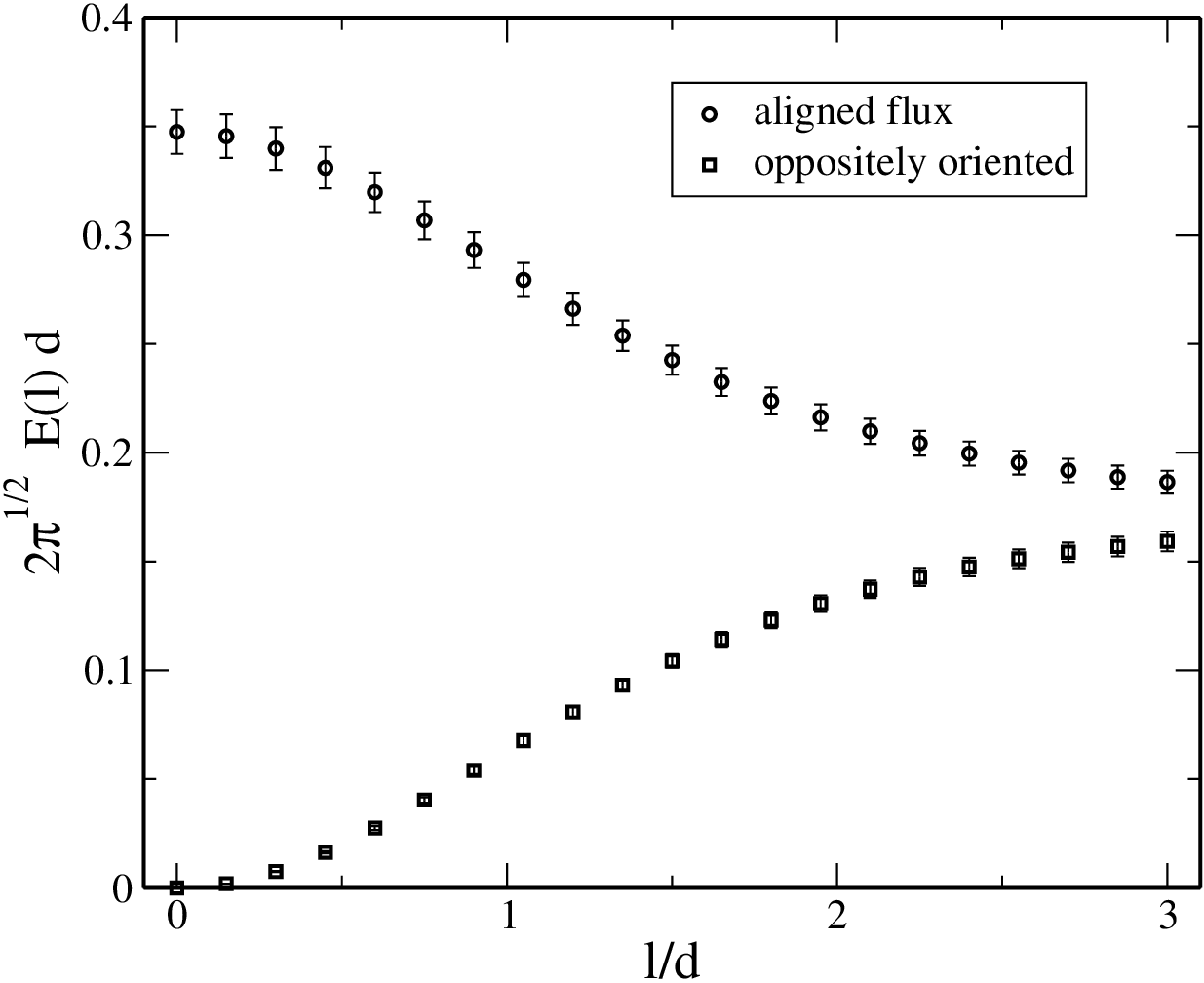} 
} 
\caption{ The interaction of two parallel vortex lines in $D=2+1$, 
$\varphi=1$, $m^2d^2=1$. 
}
\label{fig:int}
\end{figure}

\vskip 0.0cm Figure \ref{fig:cont} shows the lines of equal effective
Lagrangian $L_\mathrm{eff}^{(3)}(x,y)$ in the $xy$-plane which is
perpendicular to the vortex fluxes. The vortices are located at the
$x$ axis at a distance $l=3d$. It is straightforward (but computer
time consuming) to integrate the effective action over the $xy$-plane
in order to derive the quantum energy $E$ (\ref{eq:1.20}) of the
binary-vortex configuration. The result is shown in Fig.~\ref{fig:int}
for the case $D=2+1$. For large distances $l\gg d$, the quantum energy
approaches twice the value of a single vortex.  For $l=0$, the
vortices fall on top of each other. If the fluxes of the vortices are
oppositely oriented, the vortices annihilate each other, and the
quantum energy of the configuration vanishes. If the vortex fluxes are
equally oriented, the configuration is equivalent to the single vortex
configuration with flux $\varphi =2$. Since the quantum energy is
roughly proportional to $\varphi ^2$ (see Fig.~\ref{fig:S3m}), the
vortex configuration with flux $\varphi =2$ possesses a higher energy
than twice the energy of a single vortex, carrying flux $\varphi=1$.
Hence, vortices with an equal flux orientation repel each other in
$D=2+1$, while vortices with oppositely oriented flux attract each
other.  The same line of argument applies to the case $D=3+1$. Since
the fermionic contribution to the effective action is negative in this
case, the fermion-induced force is attractive (repulsive) for equally
(oppositely) oriented vortices, contrary to the $D=2+1$ case.

%
%

\section{Fermion condensation in a vortex gas} 
\label{sec:cond}

The fermion condensate can be related to the effective action $\Gamma
_\mathrm{eff}$ (\ref{eq:1.16}) by differentiation with respect to the
fermion mass $m$,
\begin{equation}
\int d^Dx_0\, \langle \bar{\psi} \psi \rangle \; = \; -\frac{ \partial \Gamma
_\mathrm{eff} }{ \partial m }.
\label{6.1} 
\end{equation}
Let us concentrate on $D=3+1$ dimensions in the following. Here the
condensate reads in worldline representation:
\begin{eqnarray} 
m \int d^4 x_0 \; \langle \bar{\psi} \psi \rangle &=&
 \frac{ m^2 }{4\pi^{2}} \int d^4x_0 
\int_{0}^\infty  \frac{dT}{T^2} \; \E^{-m^2 T} \;  
\biggl\langle W_{\text{spin}}[A] \, - \, 1  \biggr\rangle_x , 
\label{eq:6.2} 
\end{eqnarray}
where we have first performed the mass differentiation at an arbitrary
renormalization point and then implicitly chosen on-shell
renormalization. 

\vskip0.0cm In the present section, we intend to estimate the quark
condensate which is generated by a dilute gas of vortices of flux
$\varphi=1$ in $D=3+1$. For this purpose, we first calculate the
contribution of the single-vortex configuration with the help of
Eq.~(\ref{eq:6.2}).  Due to translation invariance of our
single-vortex background, the string tension $\chi $ (\ref{eq:1.21})
is provided in units of the vortex core size only. This allows us to
define the dimensionless function $c_0$ by
\begin{equation}
m \int d^4 x_0 \; \langle \bar{\psi} \psi \rangle_{(1)} \; = \; 
 L_t \, L_z \frac{ m^2 }{ 4\pi ^2 } \; \; 
c_0 \Bigl( m^2 \, d^2 \Bigr) \; , 
\label{eq:6.3} 
\end{equation}
where $c_0$ can directly be obtained from worldline numerics,
\begin{equation}
c_0(m^2 d^2)=2\pi \int_0^\infty
d\left(\!\frac{\rho}{d}\!\right)\,\,\frac{\rho}{d}\; 
\int _0^\infty \frac{ d\hat{T} }{\hat{T}^2} \; 
\E^{-(md)^2 \hat{T} } \,
\bigg\langle 
W_{\text{spin}}[A\,d]-1\bigg\rangle_x. \label{eq:6.4}
\end{equation}
Here all dimensionful quantities are scaled in units of the core size
$d$, e.g., $\hat{T}=T/d^2$, 
and $\rho=\sqrt{x^2+y^2}$ measures the radial distance from the
vortex core. In the following, we are interested in a dilute gas of
vortices which are static and aligned in the $z$ direction, but
intersect the $xy$ plane at random locations with random fluxes
$\varphi=\pm 1$. In this plane the vortex
gas can be characterized by a planar vortex area density
$\rho_{\text{V}}$, which is the average number of vortices per $xy$
unit area. The total number of vortices within the 4-dimensional 
spacetime is given by $N_{\text{V}} = \rho_{\text{V}} \, L_x \, L_y
$. We expect the dilute-gas approximation to give reasonable results,
if the average distance between two neighboring vortices is at least
$\gtrsim 2 d$, where the fermion-induced vortex interactions become
small (see, e.g., Fig.~\ref{fig:int} for the $D=2+1$ dimensional
analogue). In other words, there should be less than one vortex per
core-size area, $\rho_{\text{V}}\, (\pi d^2)\lesssim 1$. 
In this dilute-gas approximation, the fermion condensate averaged over
spacetime volume $V$ is therefore given by 
\begin{equation}
\langle \bar{\psi} \psi \rangle \; 
\approx \; \frac{1}{V} \, N_{\text{V}} \int d^4 x_0 \; 
  \langle \bar{\psi} \psi \rangle_{(1)}  
\; = \; \frac{ 1  }{ 4 \pi^2 } \; m \; \rho_{\text{V}} \; 
c_0 \Bigl( m^2 \, d^2 \Bigr) \; .
\label{eq:6.5} 
\end{equation}
In the limit of large fermion masses $(m \gg 1/d)$, Eq.~(\ref{eq:6.4})
can be studied analytically with the aid of the heat-kernel
expansion\re{eq:1.6}, and we find a contribution to the condensate
which is subleading in $1/m$: 
\begin{equation}
\langle \bar{\psi} \psi \rangle \; \approx \;  \frac{1}{24 \pi ^2 } \; 
\frac{1}{m} \; \frac{1}{V} \, 
\int d^4 x_0 \; F^2 (x_0) \; + \; {\cal O} \biggl( 1/m^2 \biggr) \; . 
\label{eq:6.7} 
\end{equation}
Here, we recover the familiar result~\cite{shi79} that in the
large-$m$ limit the fermion condensate is proportional to the
field strength squared of the background field 
(i.e., ``gluon condensate'' in a QCD-like language).  

\vskip0.0cm We investigate numerically the interesting regime of small
masses $m$, where the heat-kernel expansion breaks down. The
representation\re{eq:6.4} is highly convenient for this purpose.
However, as already pointed out, the loop average $\langle \ldots
\rangle $ in Eq.~(\ref{eq:6.3}) is plagued from a severe cancellation
problem which in the present case of the loop parameters employed here
limits the region of validity to $T/d^2 \lesssim 100$. In order to
ensure this limit, we confine ourselves to the mass range $m^2 T_{max}
\ge 10$, i.e., $ m^2 > 0.1 /d^2$. The investigation of the small mass
regime $(m^2 < 0.1 / d^2)$ clearly needs further study. Work
in this direction is in progress~\cite{moy02}.

\begin{figure}[t]
\centerline{  
\epsfxsize=7cm 
\epsffile{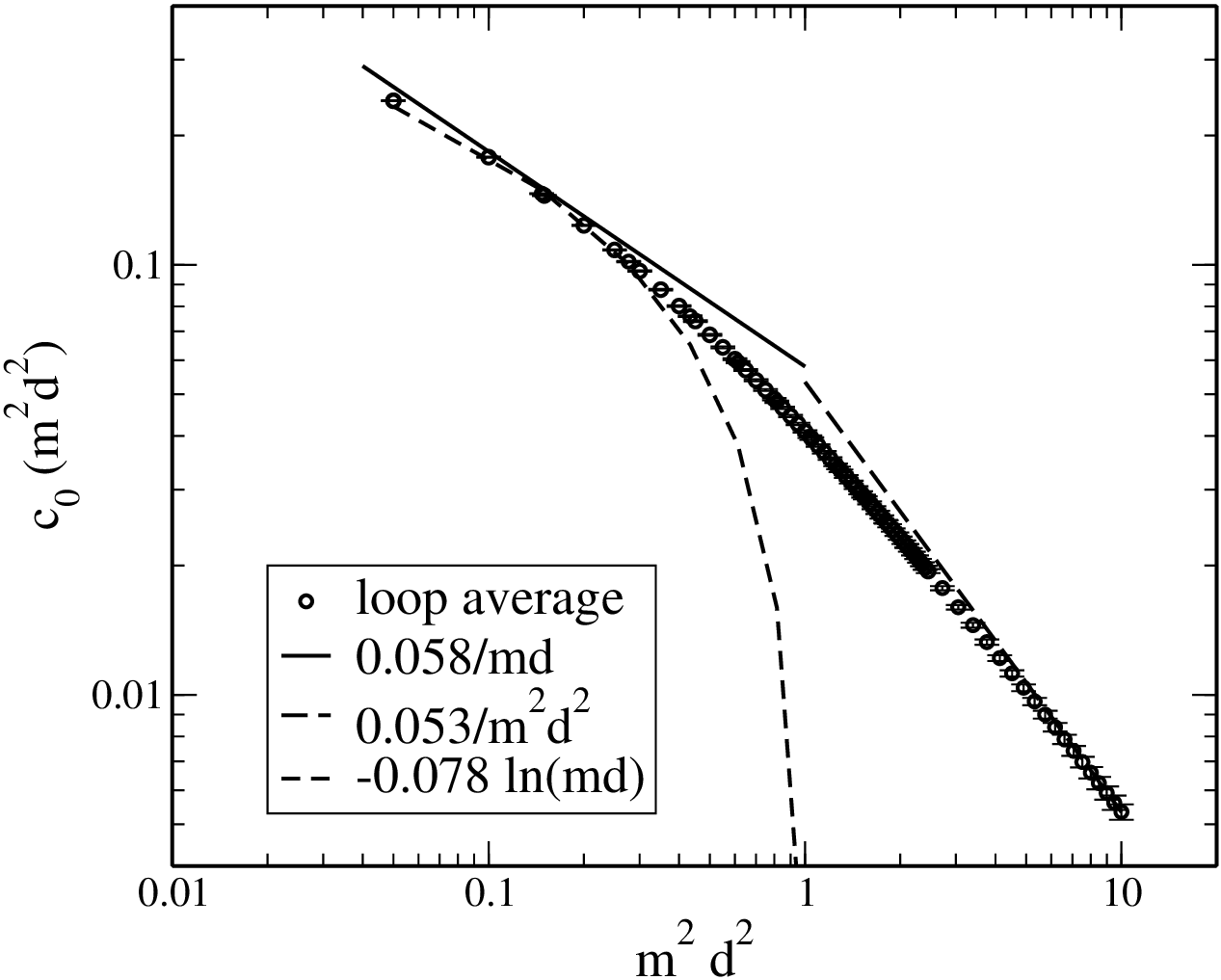} 
\hspace{0.5cm}
\epsfxsize=7cm 
\epsffile{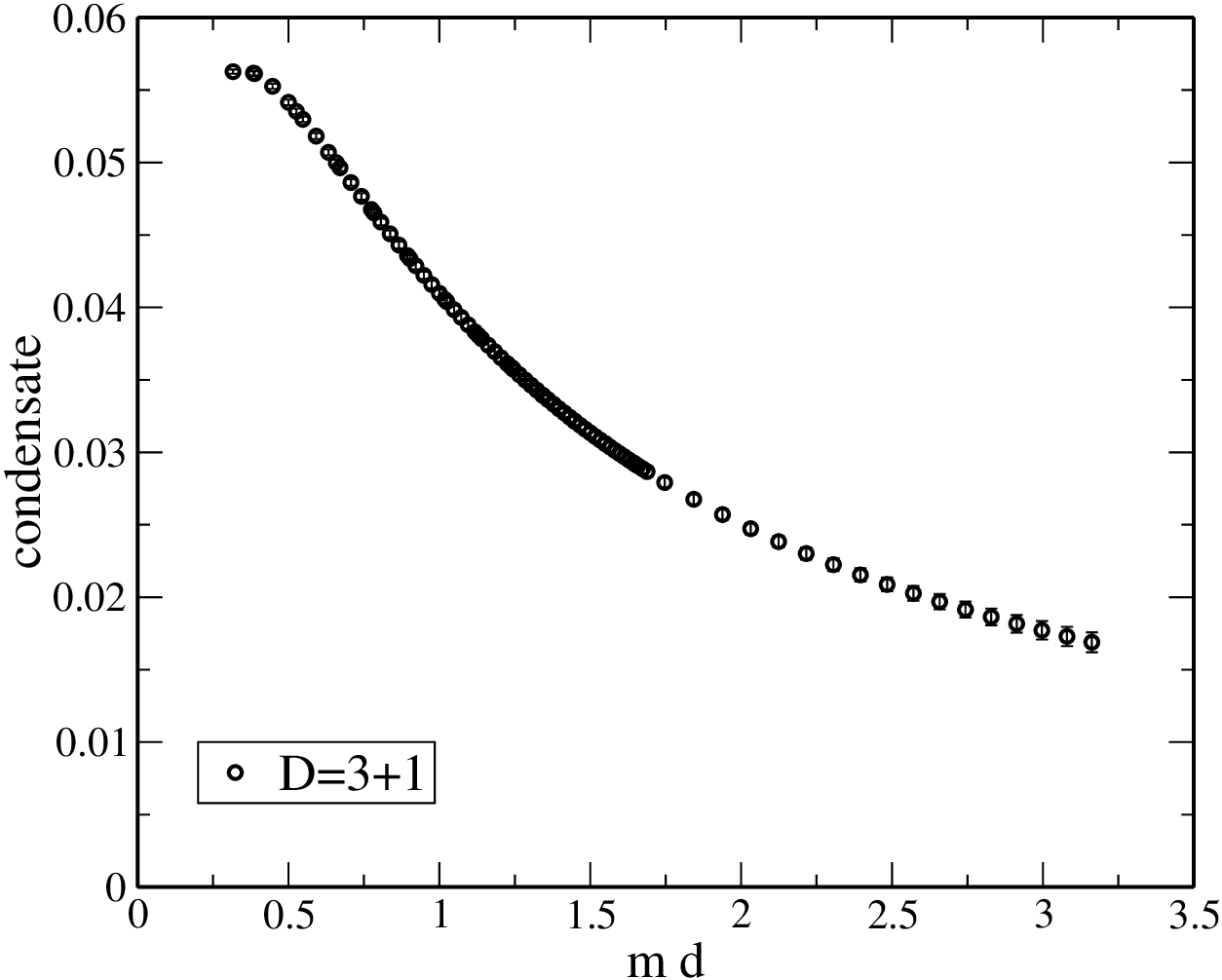} 
}
\caption{ $c_0$ (\protect{\ref{eq:6.3}}) (left panel) and the fermion condensate 
   (right panel) as function of the fermion mass $m$, 
   $\varphi$=1, $D$=3+1.}
\label{fig:cond}
\end{figure}

\vskip 0.0cm Our numerical result for $c_0$ is presented in
Fig.~\ref{fig:cond}. For small masses, i.e., $0.1 < m^2 d^2 < 0.5$, we
find that $c_0(m^2d^2)\sim 1/(md)$, which indicates that the condensate
approaches a plateau in this regime according to Eq.\re{eq:6.5},
\begin{equation}
\langle \bar{\psi} \psi \rangle \Big|_{m^2d^2={\cal O}(0.1)}
\; \approx \; \frac{1}{ 4 \pi^2 } \; m \; \rho \; \; 
c_0 \Bigl( m^2 \, d^2 \Bigr)\Big|_{m^2d^2={\cal O}(0.1)} \; \approx \; 
 \; 0.058 \; \frac{1}{ 4 \pi^2 } \; \rho \, / \, d \; ,
\label{eq:6.10} 
\end{equation}
as depicted in Fig.~\ref{fig:cond} (right panel). However, we do not
believe that this result extends to the chiral limit $m\to 0$. 
Guided by the case of a constant background field in
$D=3+1$ \cite{Dittrich:yi}, we expect that the chiral condensate
vanishes according to 
$\langle \bar{\psi} \psi \rangle \to m \ln m$ for $m\to 0$. 
As seen from Fig.~\ref{fig:cond}, our numerical result for $c_0$ 
does not discriminate between the $\ln m $ and the desired $1/m$ 
behavior in the small-mass regime. 

\vskip0.0cm Let us stress the two main findings of this subsection:
first, the vortex-gas-induced condensate is characterized by a
comparably low value of $c_0$ for all mass values depicted above and
an onset of the plateau value for comparably large masses
$m^2d^2\sim0.1$; we point out that it is not possible to fit the form
of the condensate depicted above within a derivative expansion even
qualitatively by arbitrarily varying $B$. Second, in the context of a
center vortex model for low-energy QCD, the current quark masses are
in fact finite, though small. The present result indicates that
contributions to the chiral condensate have to be expected from an
interaction of the massive 
quarks with the gauge vacuum (modeled by a random vortex background).
As a first estimate, we may insert parameter values known from lattice
calculations. In the case of an $SU(2)$ gauge theory, the density of
the center vortices is roughly given by $\rho_{\text{V}} = 3.6 /
\mathrm{fm}^2 $. The planar vortex correlation function was also
studied in SU(2) lattice gauge theory~\cite{Engelhardt:1998wu}. One
finds an exponential decrease of the correlation function. This allows
for a definition of a vortex thickness of $d\approx 0.3 \ldots 0.4
\,$fm. Since $\rho_{\text{V}}\, \pi d^2={\cal O}(0.1)$ in this case,
the dilute-gas approximation should be applicable. From
Eq.\re{eq:6.10}, we find for the vortex-induced condensate $\langle
\bar{q} q \rangle \; \approx \; [50 \, \mathrm{MeV}]^3$, for masses
$m^2d^2=0.1$, i.e., $m\simeq100$ MeV.  As discussed above, the present
algorithm cannot treat the case of light mass values, $m\simeq5\dots
10$ MeV.

\vskip 0.0cm 
The vortices considered here obviously possess a trivial topology,
i.e., they do not contribute to the topological charge. The origin of
a chiral condensate here is entirely relegated to the field
strength carried by the vortex cores.  
In a fully non-Abelian QCD-like context, topologically non-trivial vortex
configurations (due to the presence of low-lying fermionic modes of the
Dirac operator) as well as multi-gluon exchange interactions can be
expected to provide for a drastic enhancement of the condensate.

\section{Conclusions} 

The fermion-induced quantum action of Abelian vortex configurations
has been studied in the case of $D=2+1$ and $D=3+1$ dimensions using
worldline numerics~\cite{Gies:2001zp,Gies:2001tj,Schmidt:2002yd}. 
The quantum action of a single-vortex configuration is characterized
by the fermion mass $m$, the vortex thickness $d$ and the flux
$\varphi $ carried by the vortex.  Our numerical approach has been
successfully tested in the parameter regime where the derivative
expansion is expected to provide reliable results: in the
large-mass regime $m d \gg 1$ or for strong-field suppression of the
inhomogeneities $\partial^2/B \ll 1$.

\vskip 0.0cm Compared with our numerical results in $D=2+1$, the
derivative expansion provides for a reasonable approximation to the
quantum energy of a single vortex configuration even for smaller
masses, $m d \stackrel{<}{_\sim } 1$. The next-to-leading order only
adds a small correction to the leading-order result. By contrast, in
$D=3+1$, the derivative expansion is insufficient even for comparably
large masses, $md = {\cal O}(5)$. Only for very large masses, $m d\gg
1$, can the derivative expansion be trusted. We have argued that this
disparity between $D=2+1$ and $D=3+1$ arises from renormalization
effects in the latter case, which can occur near regions where the
background varies rapidly. In particular, these renormalization
effects can lead to an effective enhancement of the quantum action.
Further investigations are planned to settle this issue.

\vskip 0.0cm 
From a physical point of view, we find that the quantum action is
positive in $D=2+1$, implying that the presence of vortices is 
suppressed, and large vortex core sizes $d$ are preferred. In 
$D=3+1$, the properly renormalized quantum action turns out to be
negative: the nucleation of thin ($d \rightarrow 0$) 
vortices in $D=3+1$ is supported 
by the fermion induced quantum action. 

\vskip 0.0cm Subsequently, the binary-vortex interaction was
investigated. We found that the fermion-induced interaction favors
vortices with an opposite orientation of the fluxes in $D=2+1$, while
in $D=3+1$, a unique orientation of the fluxes is preferred.

\vskip 0.0cm It should be stressed that these statements refer to and
are derived from the fermion-induced action. In a pure QED context,
the classical action has to be taken into account. The latter will
dominate the fermion-induced action by far at weak coupling,
reflecting the usual hierarchy between classical and quantum-induced
nonlinear electrodynamics. Formally, the classical and quantum action
can become comparable in magnitude for exponentially strong fields in
the vortex core (exponentially small $d$ for fixed flux); however,
this is nothing but a manifestation of the Landau pole of QED and
thus should be rated as unphysical. 

\vskip 0.0cm Finally, the fermion condensate which is generated by a
dilute gas of vortices was studied as function of the fermion mass $md
> 0.1 $ in $D=3+1$. We found that the condensate decreases like $1/m$
for large values of the mass and that it reaches a plateau for $md
\stackrel{<}{_\sim } 0.5$. Similarly to the case of a constant
magnetic field, we expect that the condensate vanishes like $m \, \ln
m $ in the chiral limit $m \rightarrow 0$. However, our results point
to an interesting phenomenon: although chiral symmetry breaking might
be tied to topological properties of the background fields not covered
by the present considerations, the field
strength carried by the vortex cores enhances the chiral
condensate in the intermediate fermion mass regime. In a center vortex
model of low-energy QCD, this effect may lead to a nonnegligible
contribution to the condensate. We regret that our worldline
numerical algorithm in its present form cannot address the
small-mass regime $md < 0.1$ due to severe cancellations.  The study
of the chiral limit in general and chiral symmetry breaking by vortex
background fields in particular requires improved algorithms and is
left to future work~\cite{moy02}.

\vskip 1cm

\noindent 
{\bf Acknowledgments: } 

\noindent
The authors are grateful to W. Dittrich, H.  Reinhardt and H. Weigel 
for helpful discussions and detailed comments on the manuscript.
H.G.~acknowledges the support of the Deutsche Forschungsgemeinschaft
under contract Gi 328/1-1. L.M.~was supported by the Deutsche
Forschungsgemeinschaft under contract GRK683.

\end{document}